\documentclass[aip,reprint,graphicx,twocolumn,superscriptaddress,nofootinbib]{revtex4-2}
\usepackage{amssymb,amsmath,amsbsy,graphicx,xcolor,times,subfigure,marvosym,color,bm,multirow,epstopdf}

\begin{document}


\title{Magneto Optical Sensing beyond the Shot Noise Limit} 



\author{Yun-Yi Pai}
\email[]{yunyip@ornl.gov}
\affiliation{Materials Science and Technology Division, Oak Ridge National Laboratory, Oak Ridge TN 378}
\affiliation{Quantum Science Center, Oak Ridge, TN 37831 US}
\author{Claire E. Marvinney}
\affiliation{Materials Science and Technology Division, Oak Ridge National Laboratory, Oak Ridge TN 378}
\author{Chengyun Hua}
\affiliation{Materials Science and Technology Division, Oak Ridge National Laboratory, Oak Ridge TN 378}
\affiliation{Quantum Science Center, Oak Ridge, TN 37831 US}
\author{Raphael C. Pooser}
\affiliation{Computational Sciences and Engineering Division, Oak Ridge National Laboratory, Oak Ridge TN 378}
\affiliation{Quantum Science Center, Oak Ridge, TN 37831 US}
\author{Benjamin J. Lawrie}
\email[]{lawriebj@ornl.gov; This manuscript has been authored by UT-Battelle, LLC, under contract DE-AC05-00OR22725 with the US Department of Energy (DOE). The US government retains and the publisher, by accepting the article for publication, acknowledges that the US government retains a nonexclusive, paid-up, irrevocable, worldwide license to publish or reproduce the published form of this manuscript, or allow others to do so, for US government purposes. DOE will provide public access to these results of federally sponsored research in accordance with the DOE Public Access Plan (http://energy.gov/downloads/doe-public-access-plan). }
\affiliation{Materials Science and Technology Division, Oak Ridge National Laboratory, Oak Ridge TN 378}
\affiliation{Quantum Science Center, Oak Ridge, TN 37831 US}

\date{\today}

\begin{abstract}
Magneto-optical sensors including spin noise spectroscopies and magneto-optical Kerr effect microscopies are now ubiquitous tools for materials characterization that can provide new understanding of spin dynamics, hyperfine interactions, spin-orbit interactions, and charge-carrier g-factors. Both interferometric and intensity-difference measurements can provide photon shot-noise limited sensitivity, but further improvements in sensitivity with classical resources require either increased laser power that can induce unwanted heating and electronic perturbations or increased measurement times that can obscure out-of-equilibrium dynamics and radically slow experimental throughput. Proof-of-principle measurements have already demonstrated quantum enhanced spin noise measurements with a squeezed readout field that are likely to be critical to the non-perturbative characterization of spin excitations in quantum materials that emerge at low temperatures. Here, we propose a truncated nonlinear interferometric readout for low-temperature magneto-optical Kerr effect measurements that is accessible with today's quantum optical resources. We show that 10 $\text{nrad}/\sqrt{\text{Hz}}$ sensitivity is achievable with optical power as small as 1 $\mu$W such that a realistic $T$ = 83 mK can be maintained in commercially available dilution refrigerators. The quantum advantage for the proposed measurements persists even in the limit of large loss and small squeezing parameters. 
\end{abstract}


\maketitle 

\section{Introduction}
Discovered in 1844, the Faraday effect describes the circular birefringence of light transmitted through a medium. Since then, magneto-optical effects have become the cornerstone of several families of characterization techniques. Faraday and Kerr (reflection) rotation measurements are now ubiquitous tools for characterizing magnetism at microscopic to interstellar length scales \cite{jiang2013direct,RN3930}. Magneto-optical spin noise measurements have been demonstrated with broadband radio-frequency readout\cite{zapasskii2013spin,muller2010semiconductor}, and Sagnac-interferometric magneto-optical spectroscopies have been employed to discriminate various time-parity symmetries\cite{Kapitulnik_2009}. While magneto-optical spectroscopies are now a workhorse for materials characterization, state-of-the-art measurements are still limited by the photon shot noise limit (SNL). The signal to noise ratio of such shot-noise-limited measurements scales with the square root of the laser power and the readout time, but this becomes an issue when non-perturbative measurements are desired and when probing a delicate quantum mechanical state at mK temperatures \cite{lawrie2021freespace}. For instance, shot-noise-limited magneto-optical Kerr effect (MOKE) measurements offer sensitivity of order $1 \times 10^{-7} \text{rad}/\sqrt{\text{Hz}}$ with 10 $\mu W$ of laser power\cite{xia2006modified}. Resolving 10 nanoradian polarization rotation then requires either 1 mW of laser power with a 1 s measurement or 10 $\mu W$ of laser power with a 100 s measurement, or some alternative balance between increased power and increased measurement time.  Increased power will introduce substantial unwanted heating in low temperature measurements.  Longer measurement times make experiments substantially more challenging, particularly as sensitivity requirements continue to increase.

In 1981, C.M.~Caves showed that squeezed states of light enable sensitivity beyond the SNL in interferometric measurements~\cite{Caves}. Thirty years later, squeezed light is now an important resource for quantum sensing for a wide range of applications\cite{lawrie2020squeezing,lawrie2019quantum,lee2020quantum}, with examples ranging from gravitational wave detection \cite{aasi2013enhanced,ma2017proposal} to plasmonic sensing\cite{fan2015quantum,dowran2018quantum,pooser_plasmonic_2015,lee2020quantum} to scanning probe microscopies \cite{pooser_ultrasensitive_2015, pooser2020truncated, lawrie2020squeezing}, magnetometry\cite{otterstrom_nonlinear_2014,wolfgramm2010squeezed,li2018quantum,horrom2012quantum}, Raman spectroscopy\cite{michael2019squeezing,de2020quantum,casacio2021quantum}, and spin noise spectroscopy \cite{lucivero2016squeezed,wineland1992spin}. Nonlinear interferometry, in which beamsplitters are replaced with nonlinear amplifiers, has also drawn interest in recent years as an alternative to conventional quantum sensing with squeezed states~\cite{OuNatComm,Ou,jing2011realization,wang2021nonlinear}. These interferometric sensors can be designed to concentrate losses into the local oscillators, thus minimizing the detrimental effects of loss on squeezing. In addition, careful local oscillator (LO) design eliminates the need to tightly control spatial modes that proves troublesome in quantum sensors relying on multi-spatial-mode squeezed light sources~\cite{treps2003quantum,pooser_ultrasensitive_2015}. Nonlinear interferometry offers the potential to outperform classical interferometers by a factor proportional to the nonlinear gain of the amplifier~\cite{OuNatComm,Ou,jing2011realization}, and its benefit can be traced to the amount of squeezing present in the nonlinear amplifier.

Recent experiments have shown that truncated nonlinear interferometers can be used to characterize the displacement of atomic-force-microscope microcantilevers without introducing excessive heating\cite{pooser2020truncated}.  Such experiments take advantage of the squeezing generated by a nonlinear amplifier while minimizing laser induced heating by utilizing low-power or vacuum squeezed states together with high power local oscillators.  Because the shot noise level is defined by the total power at the detector, including local oscillator power, unwanted photomodification and heating of the sample can be minimized without sacrificing the quantum advantage. A growing literature now describes the sensitivity of nonlinear \cite{Ou,jing2011realization,LettPRA} and truncated nonlinear interferometers \cite{anderson2017phase,gupta2018optimized,prajapati2019polarization}.  Here, we define the fundamental limits of magneto-optical spectroscopies based on Faraday and Kerr rotation of light with classical readout fields, and we show how two-mode squeezed light can be used to obtain quantum enhanced sensitivity beyond classical limits. In a phase sensing configuration, the device corresponds to a truncated nonlinear interferometer. We characterize the effect of non-ideal homodyne detectors by incorporating a second-quantization treatment of the LOs. We find that, though the uncertainty of the polarization rotation measurement decreases fastest when both the probe and the probe LO power increase simultaneously, substantial quantum enhancement is still possible with vacuum or low power squeezed readout fields. This lays the foundation for ultra-low temperature magneto-optical spectroscopies.

\section{Polarization sensing with truncated SU(1, 1) interferometers}

\begin{figure}[htbp]
\includegraphics[width=\columnwidth]{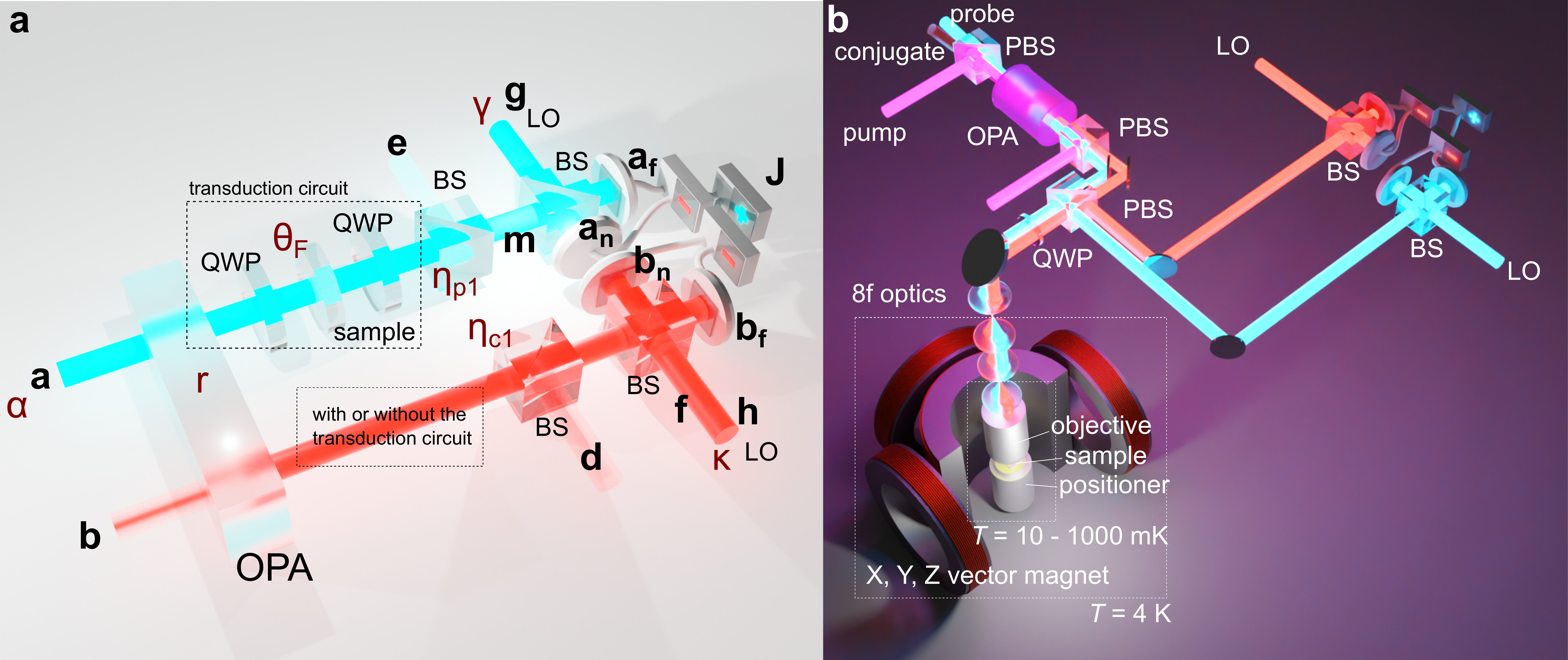}
\caption{(a) The optical setup for the calculations in this work. The boldface symbols are annihilation operators for corresponding fields. The Greek letters $\alpha$, $\gamma$, and $\kappa$ are the eigenvalues for the annihilation operators for the probe seed $\textbf{a}$, the probe LO $\textbf{g}$, and the conjugate LO $\textbf{h}$, respectively. The remaining light fields are not seeded (vacuum coupled). An optical parametric amplifier (OPA) generates a two-mode squeezed light source comprising a probe and conjugate beam that exhibit intensity difference and phase-sum squeezing. The squeezing parameter for the OPA is given by $r$. Loss in the optics train is described by beamsplitter interactions with transmission $\eta$. The final beamsplitter illustrated in each arm has $\eta=1/2$ for optimized homodyne detection. The phases for the probe and conjugate LOs are $\phi_p$ and $\phi_c$, respectively.  The homodyne detectors detect intensity differences within their two arms. The outputs of the two homodyne detectors are then summed into the joint rotated quadrature measurement operator \textbf{J}.  The optical transduction circuit is highlighted by a dashed rectangle. It consists of two quarter-wave plates that convert the Faraday or Kerr polarization rotation to a phase shift. (b) Schematic of proposed polarization rotation sensing with a truncated SU(1, 1) interferometer. The probe beam (or both probe and conjugate beams) is (are) sent into a dilution refrigerator where they are reflected off of, or transmitted through, a sample of interest resulting in a Kerr or Faraday rotation of the probe and conjugate polarization. After transducing the polarization rotation into a phase shift with a quarter-wave plate, the probe and conjugate phases are readout by dual homodyne detection. }
\label{fig:sch}
\end{figure}

The squeezed MOKE described here centers on truncated SU(1, 1) interferometry (tSU(1, 1)), wherein we transduce the MOKE polarization rotation on a two-mode squeezed state to an optical phase shift and we readout that change in phase with dual homodyne detection. However, we note that this by no means the only approach. Quantum-enhanced polarization measurements with polarization-squeezed states of light have been carried out by Lucivero et al. for instance \cite{lucivero2016squeezed}. By transducing the polarization rotation to an optical phase shift, the signal becomes immune to polarization fluctuations in optics downstream before the detector. This is particularly beneficial in implementations utilizing free-space optical components that are polarization sensitive.  For instance dielectric mirrors that help to minimize the loss in the optics train induce unwanted polarization rotation that is hard to correct for in the absence of the polarization to phase transduction described here and utilized in existing classical MOKE Sagnac interferometers\cite{xia2006modified}.

Figure \ref{fig:sch} shows the proposed optical circuit. The inset illustrates a schematic for the MOKE tSU(1, 1) interferometer considered in this work. The interferometer consist of an optical parametric amplifier (OPA) that generates a two mode squeezed state, a MOKE microscope that induces a polarization rotation on the probe and conjugate beams generated by the OPA, a quarter-wave plate that converts that polarization rotation to a phase shift, and dual homodyne detectors that read out that phase shift. The OPA is characterized by a squeezing parameter $r$ (gain $\text{G} = \text{cosh}^2(r)$). The \textbf{a}, \textbf{b}, \textbf{c}, \textbf{d}, \textbf{e}, \textbf{f}, \textbf{g}, \textbf{h}, $\mathbf{a_f}$, $\mathbf{a_n}$, $\mathbf{b_f}$, and $\mathbf{b_n}$ are optical field operators, with \textbf{a}, \textbf{b} serving as the two OPA input modes. After the probe (conjugate) is mixed with the LO \textbf{g} (\textbf{h}), the modes $\mathbf{a_f}$  and $\mathbf{a_n}$ ($\mathbf{b_f}$ and $\mathbf{b_n}$) are detected by a balanced photodetector. Beamsplitter interactions are included with efficiency $\eta_{p1}$ and $\eta_{c1}$ to allow for optical losses $1-\eta_{p1}$ and $1-\eta_{c1}$ in the probe and conjugate arms respectively and to allow for excess noise \textbf{c} and \textbf{d} to be coupled into each arm.

\begin{figure}[htbp]
\centering\includegraphics[width=\columnwidth]{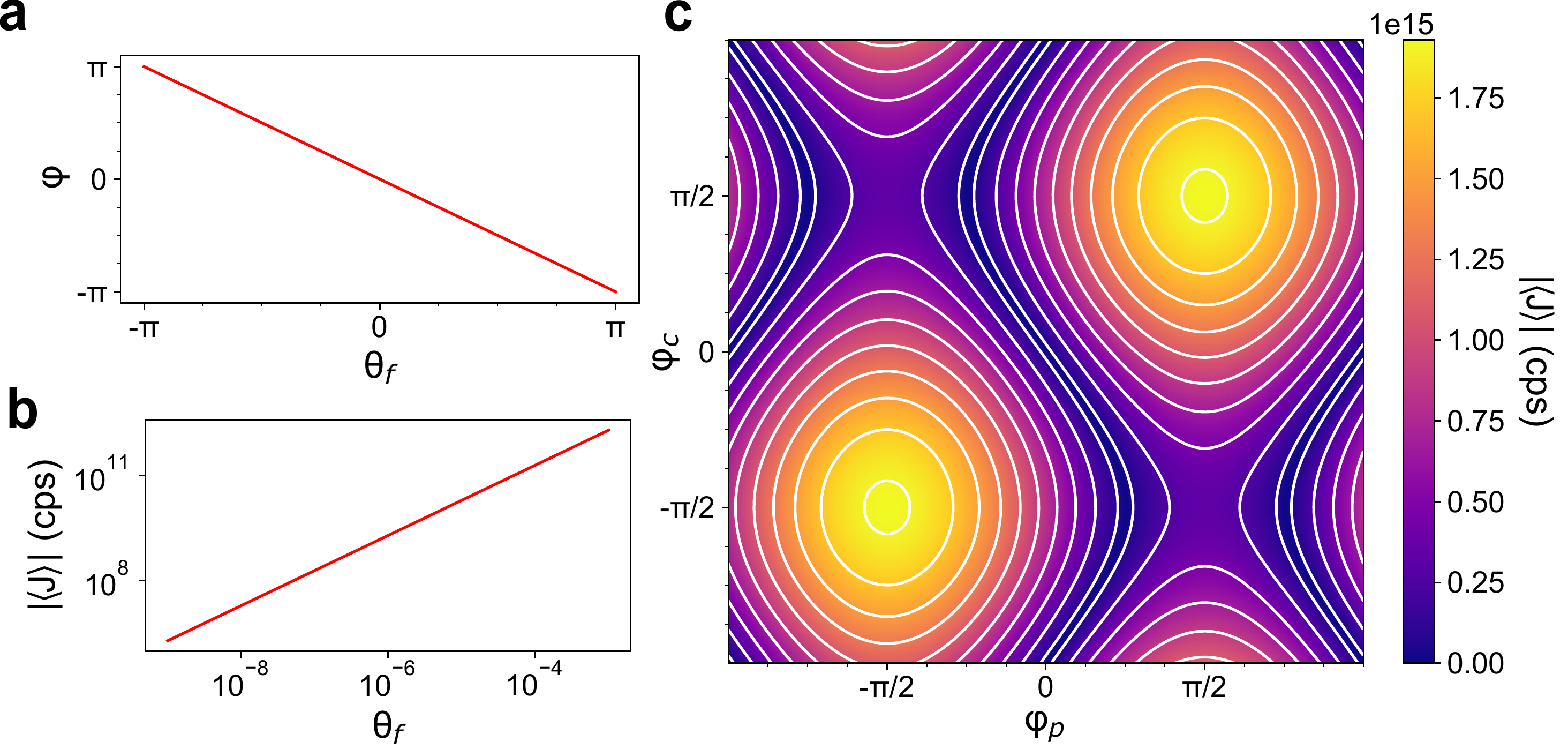}
\caption{(a) The polarization rotation to phase shift transduction is 1:1 with dual quarter-wave plates. (b) The expectation value of the joint rotated quadrature measurement operator \textbf{J} as a function of the polarization rotation $\theta_f$. (c) The expectation value of \textbf{J} as a function of the phases of the probe arm LO $\phi_p$ and conjugate arm LO $\phi_c$. Parameters for generating this plot: $r = 0.88$, $\alpha = 2\times 10^6$, $\eta = 1$, $\theta_f = 0.001$ (rad), $\gamma = \kappa = 2 \times 10^8$.  }
\label{fig:signal}
\end{figure}

The polarization rotation can be transduced into a phase shift with two quarter-wave plates (QWPs) sandwiching the sample of interest. The first QWP converts the linearly polarized light field into circularly polarized light. Due to the Faraday effect in the sample, linearly polarized light acquires a polarization rotation $\theta_F$ while circularly polarized light field acquires a phase shift. The second QWP then converts the circularly polarized light back to linearly polarized light. The resulting phase shift of the light after this optical circuit is $\phi = arg(\text{cos}(\theta_F) - i\text{sin}(\theta_F))$, where $arg(x)$ is the phase for an $x \in \mathbb{C}$. This results in a 1:1 transduction of the polarization rotation into a phase shift. Since for any optical element that can be described by a member $M$ of SU(2), and  $\forall  M \in \text{SU(2)}$, $M$ can be described by a product of two QWPs and a half wave plate (HWP) \cite{SimonMukundaMinimal}, the three real degrees of freedom for $M$ can be separately transduced to a phase shift. Though we focus on polarization rotation in cryogenic MOKE microscopies here, the same result can be extended to measurements of the ellipticity and circularity of light. 

For the probe homodyne detector, the two outputs can be expressed as 
\begin{equation}
\begin{aligned}
&\textbf{a$_f$} = \frac{\textbf{g}_{\text{LO}} \text{e}^{i \phi_p} + i \textbf{m}}{\sqrt{2}} \\
&\textbf{a$_n$} = \frac{\textbf{m} + i \textbf{g}_{\text{LO}} \text{e}^{i \phi_p}}{\sqrt{2}},
\end{aligned}
\label{eq:homodyne}
\end{equation}
where \textbf{m} is the input field into the homodyne-detector beamsplitter coming from the sample. The intensity difference measurement can be written as \textbf{a$_f$}$^\dagger$\textbf{a$_f$} - \textbf{a$_n$}$^\dagger$\textbf{a$_n$}. A similar expression can be written for the conjugate arm. With the optical circuit shown in Figure \ref{fig:sch} and with the simplification $\eta_{p1} = \eta_{c1} = \eta$, the resulting joint rotated quadrature measurement operator $J$ is: 

\begin{equation}
\begin{split}
\scriptstyle J = i \sqrt{\eta } \sinh (r) e^{i \phi_c -i \phi } h\cdot a-i \sqrt{\eta } \cosh (r) e^{i \phi -i \phi_p } g^{\dagger }\cdot a\\
\scriptstyle+ i \sqrt{\eta } \sinh (r) e^{i \phi_p -i \phi } g\cdot b-i \sqrt{\eta } \cosh (r) e^{i \phi -i \phi_c } h^{\dagger }\cdot b\\ \scriptstyle+ i \sqrt{\eta } \sinh (r) e^{i \phi_p -i \phi } g\cdot b-i \sqrt{\eta } \cosh (r) e^{i \phi -i \phi_c } h^{\dagger }\cdot b \\\scriptstyle
+ \sqrt{1-\eta } e^{-i \phi_p } g^{\dagger }\cdot c+\sqrt{1-\eta } e^{-i \phi_c } h^{\dagger }\cdot d\\\scriptstyle
+ i \sqrt{\eta } \cosh (r) e^{i \phi_p -i \phi } a^{\dagger }\cdot g+\sqrt{1-\eta } e^{i \phi_p } c^{\dagger }\cdot g\\\scriptstyle
+ i \sqrt{\eta } \cosh (r) e^{i \phi_c -i \phi } b^{\dagger }\cdot h+\sqrt{1-\eta } e^{i \phi_c } d^{\dagger }\cdot h\\\scriptstyle
- i \sqrt{\eta } \sinh (r) e^{i \phi -i \phi_c } h^{\dagger }\cdot a^{\dagger }-i \sqrt{\eta } \sinh (r) e^{i \phi -i \phi_p } g^{\dagger }\cdot b^{\dagger }\\
\end{split}
\end{equation}

{\setlength{\parindent}{0cm}
where $\phi = arg(\text{cos}(\theta_F) - i\text{sin}(\theta_F))$. The uncertainty of the polarization rotation measurement with respect to measurement $J$ is \cite{anderson2017phase, bachor2019guide}:}

\begin{equation}
\begin{aligned}
(\Delta^2 \theta_f)_J = \frac{\langle J^2 \rangle - \langle J \rangle^2}{ | \partial_{\theta_f} \langle J \rangle |^2 }
\end{aligned}
\end{equation}

\section{Limit of detection improvement}

\begin{figure}[htbp]
\centering\includegraphics[width=\columnwidth]{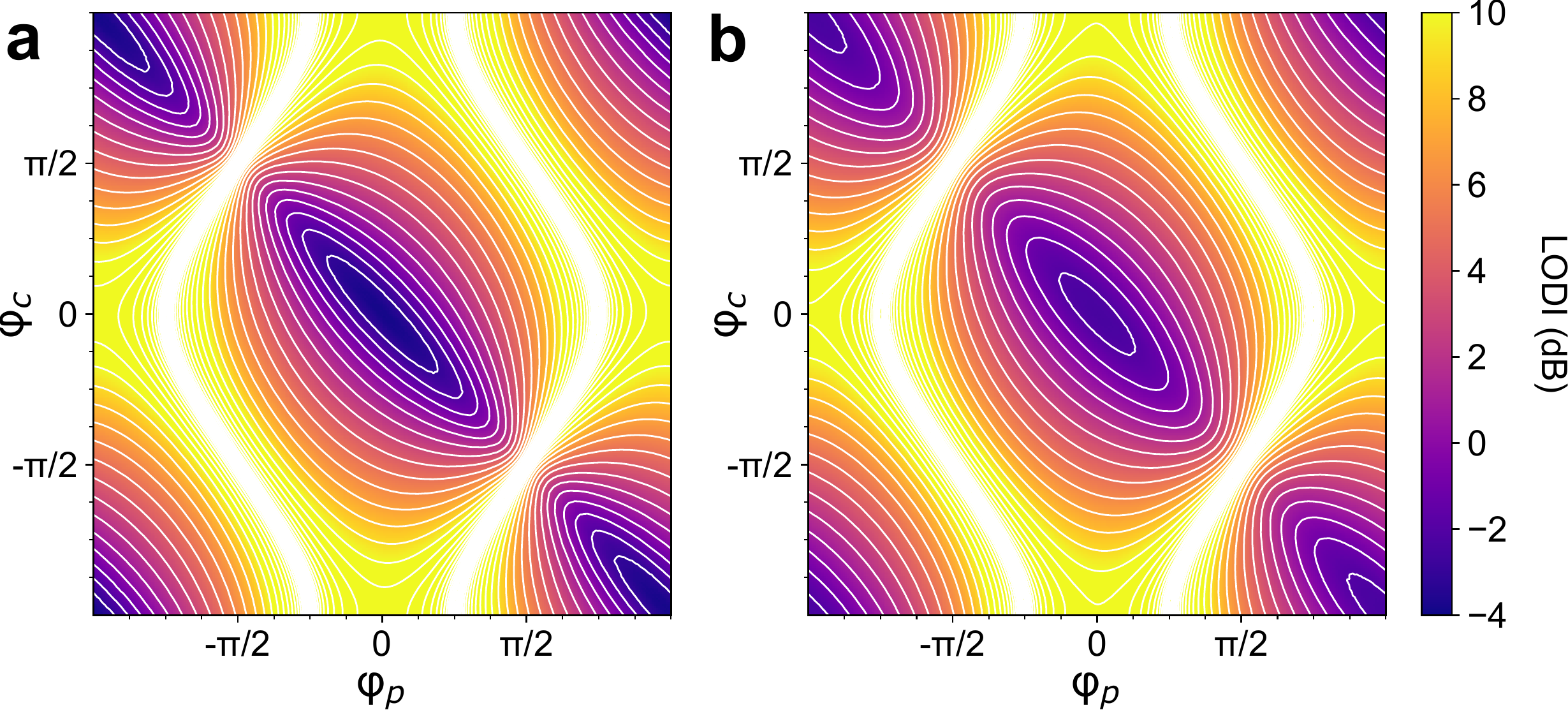}
\caption{The LODI for the tSU(1, 1) as a function of $\phi_p$ and $\phi_c$,  compared against the optimal classical interferometer.  (a) With $\alpha = 2.0 \times 10^6$,  $\gamma = \kappa = 2.0 \times 10^8$, $r$ = 0.88, $\theta_f = 0.001$ (rad), $\eta$ = 1, min(LODI) = $-$3.854 dB improvement is obtained when $\phi_p = 0.01017$ (rad), $\phi_c = 0.00117$ (rad).  (b) With $\eta$ = 0.80, the minimum of min(LODI) = $-$2.374 dB occurs at $\phi_p = - 0.02422$ (rad), $\phi_c = 0.04879$ (rad).}
\label{fig:LODI}
\end{figure}

To quantify the quantum enhancement of the signal to noise ratio, we consider a classical interferometer with both the arms interacting with the sample (Supplementary Information). The photon numbers are normalized to the photon numbers in the case of tSU(1, 1), i.e., probe arm has $|\alpha|^2 \text{cosh}^2(r)$, conjugate arm has $|\alpha|^2 \text{sinh}^2(r)$, and their LOs have $|\gamma|^2$ and $|\kappa|^2$, respectively. We define our limit of detection (LOD) equivalent to Anderson et al. \cite{anderson2017phase} :  

\begin{equation}
\text{LOD} = 10\;\text{log}_{10}\sqrt{(\Delta^2 \theta_f)_J}.
\end{equation}
The LOD can then be calculated analytically with the following starting values: $\alpha = 2.0 \times 10^6$ (or $\sim 1\mu W$ for a wavelength of 794 nm). LO $\gamma = \kappa = 2.0 \times 10^8$ (or 10 mW for the same wavelength). Initially, we define a conservative squeezing parameter of $r = 0.88$ corresponding to gain G = 3.0 dB.  With these values and only the probe photons $|\alpha|^2 \text{cosh}^2(r)$ interacting with the sample, the limit of detection improvement (LODI) ( = LOD$_{\mathrm{tSU(1, 1)}}$ $-$ LOD$_{\mathrm{classical}}$) is $- 1.50$ dB, or $- 3.78$ dB when both the probe and the conjugate interact with the sample. For high precision numeric evaluation beyond \texttt{complex256} data type, Python package \texttt{mpmath}\cite{mpmath} was used. Figure \ref{fig:LODI} (a) compares the resulting LODI for a truncated SU(1, 1) interferometer against a classical interferometer for polarization rotation sensing, as a function of $\phi_p$ and $\phi_c$, the phases of probe and conjugate LOs . The optimal setting for  $\phi_p$ and $\phi_c$ is a function of $\theta_f$ and occurs at $\phi_p = 0.01017$ (rad), $\phi_c = 0.00117$ (rad) for $\theta_f = 0.001$ (rad). When $\eta < 1$, the basin of the contour becomes more flat, as shown in Figure \ref{fig:LODI} (b) with $\eta = 0.8$ and the LODI deteriorates to $-$2.374 dB. 

\section{Dependence on Experimental Parameters}

\begin{figure}[htbp]
\centering\includegraphics[width=\columnwidth]{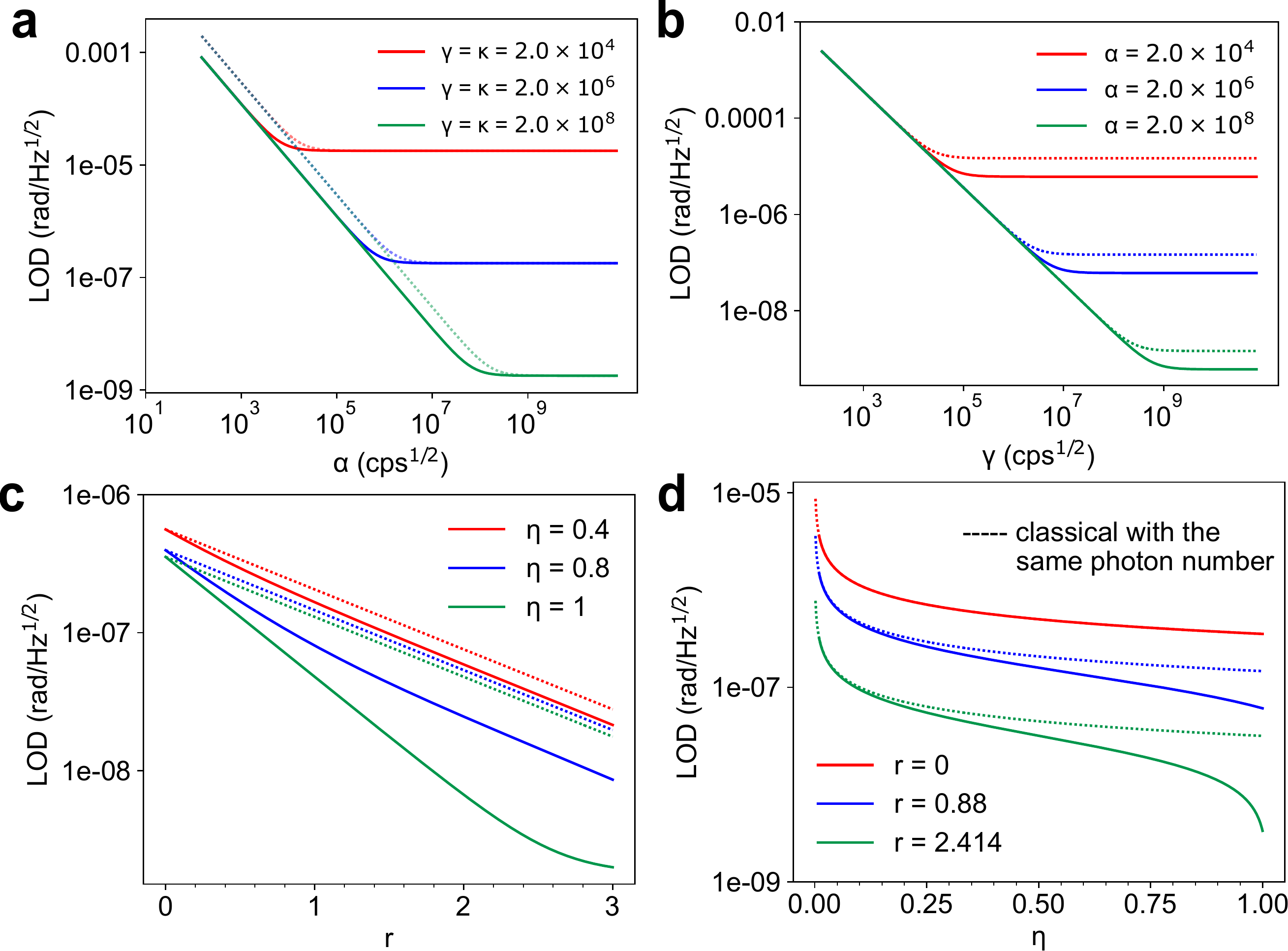}
\caption{LOD for tSU(1, 1) and classical interferometer as a function of several experimental parameters. All the parameters that are not explicitly mentioned are the set of the starting values, i.e., $\alpha = 2.0 \times 10^6$,$ \gamma = \kappa = 2.0 \times 10^8$ (with appropriate scaling for the classical interferometer scheme), $r = 0.88$, $\eta$ = 1, $\theta_f = 0.001$. (a) LOD as a function of $\alpha$ at various $\gamma = \kappa$. (b) LOD as a function of $\gamma = \kappa$ at various $\alpha$. (c) LOD as a function of $r$ at various $\eta$. (d) LOD as a function of $\eta$ at various $r$.}
\label{fig:lincuts}
\end{figure}

Figure \ref{fig:lincuts} is a collection of LOD for tSU(1, 1) and classical interferometer as a function of various experimental parameters from $\alpha$, $\gamma$, $\eta$ and $r$. Dashed plots are the classical case with the same amount of photons interacting with the sample on both the probe and the conjugate arm and the same amount of photons at the LOs. 

\begin{enumerate}
    \item $\alpha$. As shown in Figure \ref{fig:lincuts} (a), the LOD improvement appears at lower $\alpha$  and that advantage disappears as $\alpha$ approaches $\gamma$. The advantage at low $\alpha$ in fact extends all the way to $\alpha \to 0$, which is corresponding to seeding only the vacuum states into the OPA though LOD is ill-defined at $\alpha = 0$. We revisit the vacuum seeding in the the next Section.  
    \item $\gamma$. In \ref{fig:lincuts} (b), as $\gamma$ increases, the LOD starts to improve for the tSU(1, 1) over classical interferometer as $\gamma \sim \alpha$ and the slope of the improvement drastically decreases when $\gamma \sim 100 \; \alpha$ and saturates when the shot noise from the LOs dominates the detectors. 
    \item $r$. In Figure \ref{fig:lincuts} (c), the improvement in LOD starts at $r = 0$, but the margin decreases as $\eta$ decreases. 
    \item $\eta$. As in Figure \ref{fig:lincuts} (d), at $r = 0$, the LOD for the tSU(1, 1) and the classical case conincide with each other. The LOD improvement started to be visible as $\eta = 0$ as long as $r > 0$. As $r$ increases, the LOD becomes increasingly sensitive to $\eta$, especially at the $\eta \sim 1$ limit. For example, when $r = 2.414 $ (G = 15 dB) and $\eta = 0.92$, the LOD improvement is 5.29 dB. But with G = 20 dB at $r = 3.000$ and $\eta = 0.92$, the LODI is only 5.41 dB.   
\end{enumerate}

\begin{figure}[htbp]
\centering\includegraphics[width=\columnwidth]{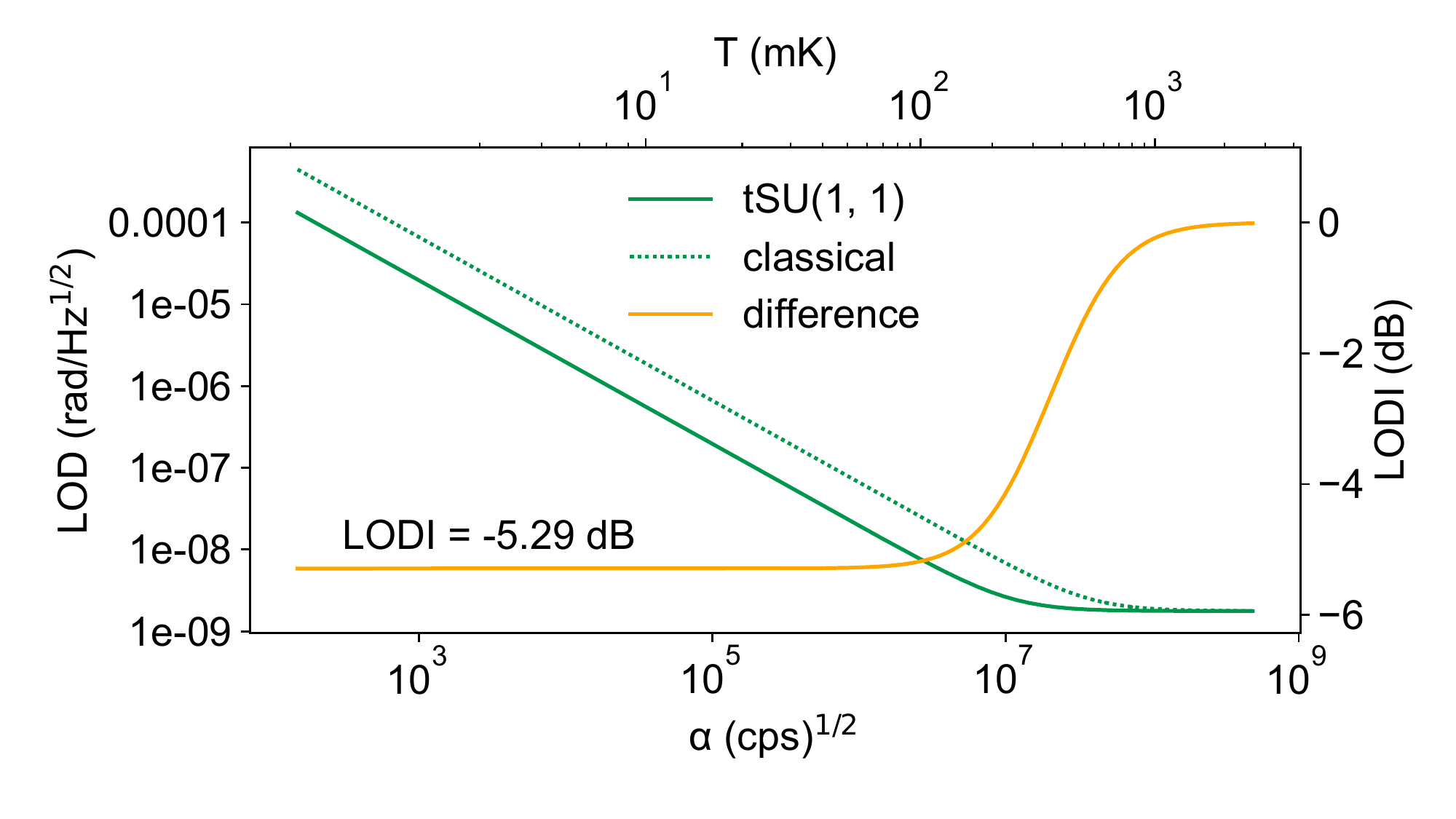}
\caption{The limit of detection improvement (LODI) for the tSU(1, 1) and classical interferometer as a function of $\alpha$ and modeled sample temperature $T$, with $\gamma = \kappa = 2.0 \times 10^{8}$, $r$ = 2.413, $\eta$ = 0.92. The dilution refrigerator with optical access for this modeling was described in Ref. \cite{lawrie2021freespace}. The system has cooling power 1 mW at $T =$ 100 mK with 200 $\mu$W of cooling power available on the cold insertable probe. The data for modeling the temperature dependence of the probe power is from the same reference.}
\label{fig:LOD_tmp}
\end{figure}

In order for magneto-optical microscopies to probe magnetic structure present in quantum materials at mK temperatures, the measurement must be non-perturbative; the laser readout should not heat the sample over a phase transition during the measurement.  In order to understand the optimal sensitivity that can be achieved for a given probe photon number at mK temperatures, we modeled the temperature as a function of the probe photon number for the dilution refrigerator described in Ref. \cite{lawrie2021freespace}. The system has cooling power of 1 mW at $T =$ 100 mK at the mixing chamber plate, 200 $\mu$W on the cold insertable probe, and less at the end of the cold finger, where the sample is mounted. Figure \ref{fig:LOD_tmp} illustrates the LOD and LODI as a function of $\alpha$ and $T$, with state-of-the-art squeezing parameter $r$ = 2.413 (G = 15 dB) and $\eta$ = 0.92 \cite{Vahlbruch15dB}. A consistent LODI of -5.29 dB extends to the low $\alpha$ limit. At $\alpha = 1.97 \times 10^{6}$, the LODI is -5.27 dB, $T$ = 83 mK, and LOD $=$ 10 $\text{nrad}/\sqrt{\text{Hz}}$. 

\section{Sensing with two mode squeezed vacuum}

\begin{figure}[htbp]
\centering\includegraphics[width=\columnwidth]{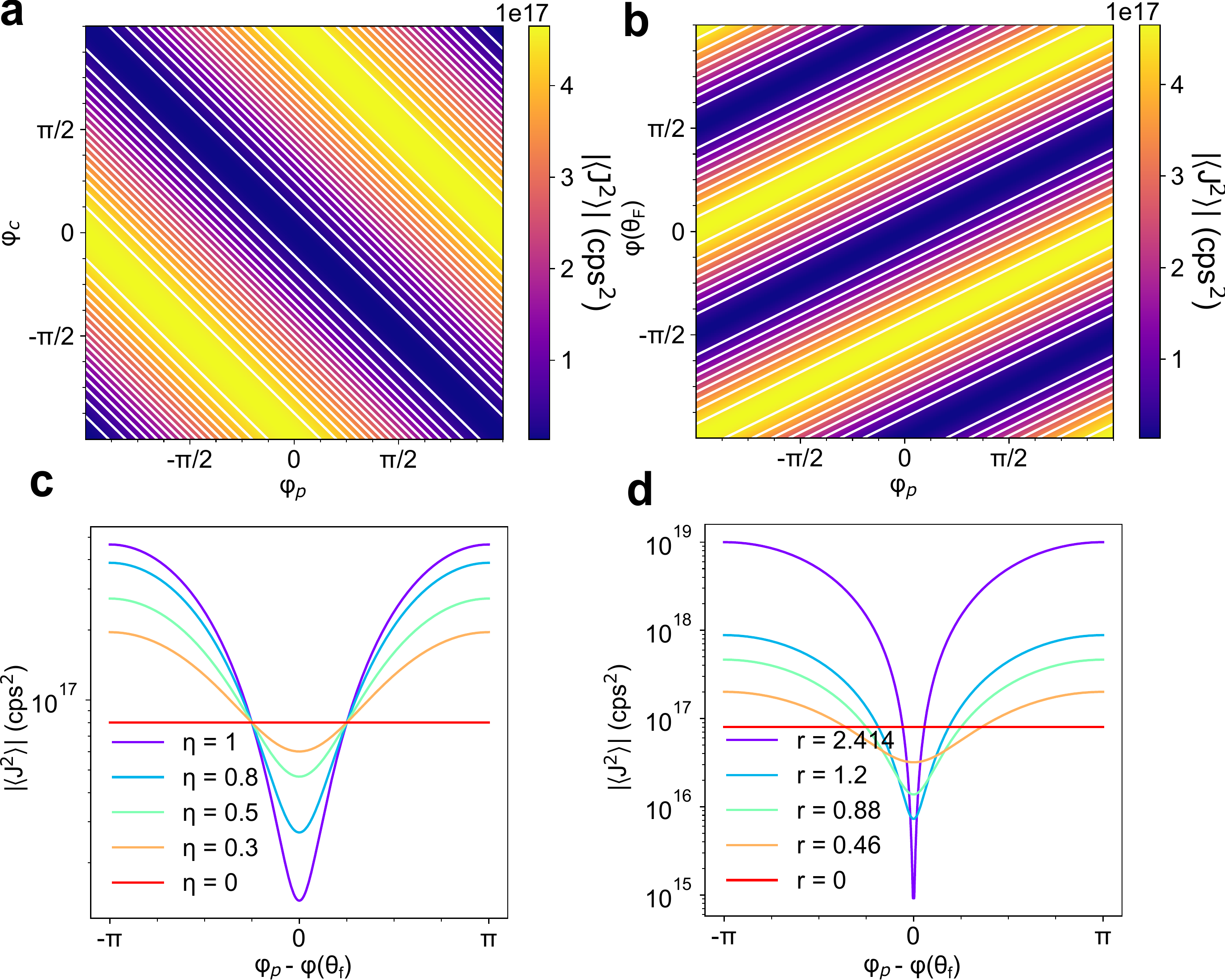}
\caption{Uncertainty $\langle J^2 \rangle$ when sensing polarization rotation with two mode squeezed vacuum. (a) $\langle J^2 \rangle$  as a function of $\phi_p$ and $\phi_c$ when $\phi (\theta_f) = 0.001$. (b) Minima of $\langle J^2 \rangle$ occurs when the $\phi (\theta_f) = \phi_p/2$. (c) $\langle J^2 \rangle$  as a function of $\phi_p - \phi (\theta_f)$ at several $\eta$  values with $r = 0.88$. (d) $\langle J^2 \rangle$  as a function of $\phi_p - \phi (\theta_f)$ at several $r$  value with $\eta = 1$.} 
\label{fig:vacuum}
\end{figure}

When neither of the two input ports of the OPA are seeded, the output of the OPA is a two mode squeezed vacuum. Though they individually are thermal states, the joint quadrature of the two exhibits quantum noise reduction as well as entanglement. It is therefore possible to be used for sensing. With the seeded photon $\alpha = 0$, the $\langle J \rangle$ vanishes as well as $| \partial_{\theta_f} \langle J \rangle |$, where as $\langle J^2 \rangle$ and hence $\Delta^2 J$ does not. Additionally, the resulting $\Delta^2 J$ is a function of $\phi$ where $\phi = \theta_f$ if using the same transduction described earlier. This is not possible for the classical case, as the resulting $\Delta^2 J$ does not depend on $\phi$ and only depends on the LO power. With one arm interacting with the sample, the  $\Delta^2 J$ is:  
\begin{multline}
\scriptstyle \Delta^2 J  = 2 \eta  \sinh (r) \left(\sinh (r)-\gamma  \kappa  \cosh (r) e^{-i (2\phi -\phi_p -\phi_c )}\right) \\ \scriptstyle 
+\gamma  \left(\gamma  (-\eta )+\gamma +\gamma  \eta  \cosh (2 r)-\eta  \kappa  \sinh (2 r) e^{i (2\phi -\phi_p -\phi_c )}\right) \\ \scriptstyle 
+\kappa ^2 (-\eta +\eta  \cosh (2 r)+1)
\end{multline}

Figure \ref{fig:vacuum} (a) represents $\Delta^2 J$ as a function of $\phi_p$ and $\phi_c$ when $\phi (\theta_f) = 0.001$, and Fig. \ref{fig:vacuum} (b) represents $\Delta^2 J$ as a function of $\phi (\theta_f)$ and $\phi_p$. $\Delta^2 J$ has minima when $\phi (\theta_f) = \phi_p/2$ and maxima when they are out of phase by $\phi (\theta_f) = \phi_p/2 + \pi/2$. Figure \ref{fig:vacuum} (c) represents $\langle J^2 \rangle$  as a function of $\phi_p - \phi (\theta_f)$ at several $\eta$  values with $r = 0.88$. Figure \ref{fig:vacuum} (d) represents $\langle J^2 \rangle$  as a function of $\phi_p - \phi (\theta_f)$ at several $r$ values with $\eta = 1$.  The quantum advantage for two-mode-squeezed-vacuum sensing exists as long as $\eta > 0$ and $r > 0$. This makes the two mode squeezed vacuum extremely desirable for quantum sensing at extremely low-temperature, low-probe-power conditions. 

\section{Conclusion}
In summary, we have described a clear path to quantum-enhanced MOKE at ultra low temperatures that is accessible with today's quantum light sources. We have derived an analytical expression for the joint rotated quadrature detection operator for polarization rotation measurements using a tSU(1, 1) interferometer. We have shown that the tSU(1, 1) outperforms its classical counterpart MZI in polarization rotation measurements across the entire parameter space explored here. We further showed that the measurement can benefit from a strong LO when the probe photon number has to be limited to minimize heating. The calculation done here is automatically generated for any sample that can be described by an SU(2) group. The calculation for the Sagnac version is expect to hold the parity etc. The calculation here can be extended to the frequency domain via the dissipation fluctuation theorem. The treatment presented here assumed a single-spatial-mode squeezed light source, but it extends trivially to multi-spatial-mode squeezed light sources, which is critical to the detection of polar and longitudinal MOKE signals on segmented quadrant photodiodes. Thus, with today's multi-spatial-mode squeezed light sources, it is possible to characterize polar and longitudinal MOKE signals with LOD $=$  10 $\text{nrad}/\sqrt{\text{Hz}}$ at temperatures of less than 100 mK.


\begin{acknowledgments}
The connections between quantum sensing theory and condensed matter physics were sponsored by the U. S. Department of Energy, Office of Science, Basic Energy Sciences, Materials Sciences and Engineering Division. The foundations of the quantum sensing theory were supported by the U.S. Department of Energy, Office of Science, National Quantum Information Science Research Centers, Quantum Science Center. Postdoctoral (CEM) research support was provided by the Intelligence Community Postdoctoral Research Fellowship Program at the Oak Ridge National Laboratory, administered by Oak Ridge Institute for Science and Education through an interagency agreement between the U.S. Department of Energy and the Office of the Director of National Intelligence. 
\end{acknowledgments}

\section*{Data Availability Statement}
Data underlying the results presented in this paper may be obtained from the authors upon reasonable request.

\bibliography{references.bib}

\begin{thebibliography}{40}%
\makeatletter
\providecommand \@ifxundefined [1]{%
 \@ifx{#1\undefined}
}%
\providecommand \@ifnum [1]{%
 \ifnum #1\expandafter \@firstoftwo
 \else \expandafter \@secondoftwo
 \fi
}%
\providecommand \@ifx [1]{%
 \ifx #1\expandafter \@firstoftwo
 \else \expandafter \@secondoftwo
 \fi
}%
\providecommand \natexlab [1]{#1}%
\providecommand \enquote  [1]{``#1''}%
\providecommand \bibnamefont  [1]{#1}%
\providecommand \bibfnamefont [1]{#1}%
\providecommand \citenamefont [1]{#1}%
\providecommand \href@noop [0]{\@secondoftwo}%
\providecommand \href [0]{\begingroup \@sanitize@url \@href}%
\providecommand \@href[1]{\@@startlink{#1}\@@href}%
\providecommand \@@href[1]{\endgroup#1\@@endlink}%
\providecommand \@sanitize@url [0]{\catcode `\\12\catcode `\$12\catcode
  `\&12\catcode `\#12\catcode `\^12\catcode `\_12\catcode `\%12\relax}%
\providecommand \@@startlink[1]{}%
\providecommand \@@endlink[0]{}%
\providecommand \url  [0]{\begingroup\@sanitize@url \@url }%
\providecommand \@url [1]{\endgroup\@href {#1}{\urlprefix }}%
\providecommand \urlprefix  [0]{URL }%
\providecommand \Eprint [0]{\href }%
\providecommand \doibase [0]{https://doi.org/}%
\providecommand \selectlanguage [0]{\@gobble}%
\providecommand \bibinfo  [0]{\@secondoftwo}%
\providecommand \bibfield  [0]{\@secondoftwo}%
\providecommand \translation [1]{[#1]}%
\providecommand \BibitemOpen [0]{}%
\providecommand \bibitemStop [0]{}%
\providecommand \bibitemNoStop [0]{.\EOS\space}%
\providecommand \EOS [0]{\spacefactor3000\relax}%
\providecommand \BibitemShut  [1]{\csname bibitem#1\endcsname}%
\let\auto@bib@innerbib\@empty
\bibitem [{\citenamefont {Jiang}\ \emph {et~al.}(2013)\citenamefont {Jiang},
  \citenamefont {Upadhyaya}, \citenamefont {Fan}, \citenamefont {Zhao},
  \citenamefont {Wang}, \citenamefont {Chang}, \citenamefont {Lang},
  \citenamefont {Wong}, \citenamefont {Lewis}, \citenamefont {Lin} \emph
  {et~al.}}]{jiang2013direct}%
  \BibitemOpen
  \bibfield  {author} {\bibinfo {author} {\bibfnamefont {W.}~\bibnamefont
  {Jiang}}, \bibinfo {author} {\bibfnamefont {P.}~\bibnamefont {Upadhyaya}},
  \bibinfo {author} {\bibfnamefont {Y.}~\bibnamefont {Fan}}, \bibinfo {author}
  {\bibfnamefont {J.}~\bibnamefont {Zhao}}, \bibinfo {author} {\bibfnamefont
  {M.}~\bibnamefont {Wang}}, \bibinfo {author} {\bibfnamefont {L.-T.}\
  \bibnamefont {Chang}}, \bibinfo {author} {\bibfnamefont {M.}~\bibnamefont
  {Lang}}, \bibinfo {author} {\bibfnamefont {K.~L.}\ \bibnamefont {Wong}},
  \bibinfo {author} {\bibfnamefont {M.}~\bibnamefont {Lewis}}, \bibinfo
  {author} {\bibfnamefont {Y.-T.}\ \bibnamefont {Lin}}, \emph {et~al.},\
  }\bibfield  {title} {\enquote {\bibinfo {title} {Direct imaging of thermally
  driven domain wall motion in magnetic insulators},}\ }\href@noop {}
  {\bibfield  {journal} {\bibinfo  {journal} {Physical Review Letters}\
  }\textbf {\bibinfo {volume} {110}},\ \bibinfo {pages} {177202} (\bibinfo
  {year} {2013})}\BibitemShut {NoStop}%
\bibitem [{\citenamefont {Gray}\ \emph {et~al.}(1998)\citenamefont {Gray},
  \citenamefont {Landecker}, \citenamefont {Dewdney},\ and\ \citenamefont
  {Taylor}}]{RN3930}%
  \BibitemOpen
  \bibfield  {author} {\bibinfo {author} {\bibfnamefont {A.~D.}\ \bibnamefont
  {Gray}}, \bibinfo {author} {\bibfnamefont {T.~L.}\ \bibnamefont {Landecker}},
  \bibinfo {author} {\bibfnamefont {P.~E.}\ \bibnamefont {Dewdney}},\ and\
  \bibinfo {author} {\bibfnamefont {A.~R.}\ \bibnamefont {Taylor}},\ }\bibfield
   {title} {\enquote {\bibinfo {title} {A large-scale, interstellar
  faraday-rotation feature of unknown origin},}\ }\href
  {https://doi.org/10.1038/31413} {\bibfield  {journal} {\bibinfo  {journal}
  {Nature}\ }\textbf {\bibinfo {volume} {393}},\ \bibinfo {pages} {660--662}
  (\bibinfo {year} {1998})}\BibitemShut {NoStop}%
\bibitem [{\citenamefont {Zapasskii}(2013)}]{zapasskii2013spin}%
  \BibitemOpen
  \bibfield  {author} {\bibinfo {author} {\bibfnamefont {V.~S.}\ \bibnamefont
  {Zapasskii}},\ }\bibfield  {title} {\enquote {\bibinfo {title} {Spin-noise
  spectroscopy: from proof of principle to applications},}\ }\href@noop {}
  {\bibfield  {journal} {\bibinfo  {journal} {Advances in Optics and
  Photonics}\ }\textbf {\bibinfo {volume} {5}},\ \bibinfo {pages} {131--168}
  (\bibinfo {year} {2013})}\BibitemShut {NoStop}%
\bibitem [{\citenamefont {M{\"u}ller}\ \emph {et~al.}(2010)\citenamefont
  {M{\"u}ller}, \citenamefont {Oestreich}, \citenamefont {R{\"o}mer},\ and\
  \citenamefont {H{\"u}bner}}]{muller2010semiconductor}%
  \BibitemOpen
  \bibfield  {author} {\bibinfo {author} {\bibfnamefont {G.~M.}\ \bibnamefont
  {M{\"u}ller}}, \bibinfo {author} {\bibfnamefont {M.}~\bibnamefont
  {Oestreich}}, \bibinfo {author} {\bibfnamefont {M.}~\bibnamefont
  {R{\"o}mer}},\ and\ \bibinfo {author} {\bibfnamefont {J.}~\bibnamefont
  {H{\"u}bner}},\ }\bibfield  {title} {\enquote {\bibinfo {title}
  {Semiconductor spin noise spectroscopy: Fundamentals, accomplishments, and
  challenges},}\ }\href@noop {} {\bibfield  {journal} {\bibinfo  {journal}
  {Physica E: Low-dimensional Systems and Nanostructures}\ }\textbf {\bibinfo
  {volume} {43}},\ \bibinfo {pages} {569--587} (\bibinfo {year}
  {2010})}\BibitemShut {NoStop}%
\bibitem [{\citenamefont {Kapitulnik}\ \emph {et~al.}(2009)\citenamefont
  {Kapitulnik}, \citenamefont {Xia}, \citenamefont {Schemm},\ and\
  \citenamefont {Palevski}}]{Kapitulnik_2009}%
  \BibitemOpen
  \bibfield  {author} {\bibinfo {author} {\bibfnamefont {A.}~\bibnamefont
  {Kapitulnik}}, \bibinfo {author} {\bibfnamefont {J.}~\bibnamefont {Xia}},
  \bibinfo {author} {\bibfnamefont {E.}~\bibnamefont {Schemm}},\ and\ \bibinfo
  {author} {\bibfnamefont {A.}~\bibnamefont {Palevski}},\ }\bibfield  {title}
  {\enquote {\bibinfo {title} {Polar kerr effect as probe for time-reversal
  symmetry breaking in unconventional superconductors},}\ }\href
  {https://doi.org/10.1088/1367-2630/11/5/055060} {\bibfield  {journal}
  {\bibinfo  {journal} {New Journal of Physics}\ }\textbf {\bibinfo {volume}
  {11}},\ \bibinfo {pages} {055060} (\bibinfo {year} {2009})}\BibitemShut
  {NoStop}%
\bibitem [{\citenamefont {Lawrie}\ \emph {et~al.}(2021)\citenamefont {Lawrie},
  \citenamefont {Feldman}, \citenamefont {Marvinney},\ and\ \citenamefont
  {Pai}}]{lawrie2021freespace}%
  \BibitemOpen
  \bibfield  {author} {\bibinfo {author} {\bibfnamefont {B.~J.}\ \bibnamefont
  {Lawrie}}, \bibinfo {author} {\bibfnamefont {M.}~\bibnamefont {Feldman}},
  \bibinfo {author} {\bibfnamefont {C.~E.}\ \bibnamefont {Marvinney}},\ and\
  \bibinfo {author} {\bibfnamefont {Y.~Y.}\ \bibnamefont {Pai}},\ }\href@noop
  {} {\enquote {\bibinfo {title} {Free-space confocal magneto-optical
  spectroscopies at millikelvin temperatures},}\ } (\bibinfo {year} {2021}),\
  \Eprint {https://arxiv.org/abs/2103.06851} {arXiv:2103.06851 [quant-ph]}
  \BibitemShut {NoStop}%
\bibitem [{\citenamefont {Xia}\ \emph {et~al.}(2006)\citenamefont {Xia},
  \citenamefont {Beyersdorf}, \citenamefont {Fejer},\ and\ \citenamefont
  {Kapitulnik}}]{xia2006modified}%
  \BibitemOpen
  \bibfield  {author} {\bibinfo {author} {\bibfnamefont {J.}~\bibnamefont
  {Xia}}, \bibinfo {author} {\bibfnamefont {P.~T.}\ \bibnamefont {Beyersdorf}},
  \bibinfo {author} {\bibfnamefont {M.~M.}\ \bibnamefont {Fejer}},\ and\
  \bibinfo {author} {\bibfnamefont {A.}~\bibnamefont {Kapitulnik}},\ }\bibfield
   {title} {\enquote {\bibinfo {title} {Modified sagnac interferometer for
  high-sensitivity magneto-optic measurements at cryogenic temperatures},}\
  }\href@noop {} {\bibfield  {journal} {\bibinfo  {journal} {Applied Physics
  Letters}\ }\textbf {\bibinfo {volume} {89}},\ \bibinfo {pages} {062508}
  (\bibinfo {year} {2006})}\BibitemShut {NoStop}%
\bibitem [{\citenamefont {Caves}(1981)}]{Caves}%
  \BibitemOpen
  \bibfield  {author} {\bibinfo {author} {\bibfnamefont {C.~M.}\ \bibnamefont
  {Caves}},\ }\bibfield  {title} {\enquote {\bibinfo {title}
  {Quantum-mechanical noise in an interferometer},}\ }\href
  {https://doi.org/10.1103/PhysRevD.23.1693} {\bibfield  {journal} {\bibinfo
  {journal} {Physical Review D}\ }\textbf {\bibinfo {volume} {23}},\ \bibinfo
  {pages} {1693--1708} (\bibinfo {year} {1981})}\BibitemShut {NoStop}%
\bibitem [{\citenamefont {Lawrie}, \citenamefont {Pooser},\ and\ \citenamefont
  {Maksymovych}(2020)}]{lawrie2020squeezing}%
  \BibitemOpen
  \bibfield  {author} {\bibinfo {author} {\bibfnamefont {B.}~\bibnamefont
  {Lawrie}}, \bibinfo {author} {\bibfnamefont {R.}~\bibnamefont {Pooser}},\
  and\ \bibinfo {author} {\bibfnamefont {P.}~\bibnamefont {Maksymovych}},\
  }\bibfield  {title} {\enquote {\bibinfo {title} {Squeezing noise in
  microscopy with quantum light},}\ }\href@noop {} {\bibfield  {journal}
  {\bibinfo  {journal} {Trends in Chemistry}\ }\textbf {\bibinfo {volume}
  {2}},\ \bibinfo {pages} {683--686} (\bibinfo {year} {2020})}\BibitemShut
  {NoStop}%
\bibitem [{\citenamefont {Lawrie}\ \emph {et~al.}(2019)\citenamefont {Lawrie},
  \citenamefont {Lett}, \citenamefont {Marino},\ and\ \citenamefont
  {Pooser}}]{lawrie2019quantum}%
  \BibitemOpen
  \bibfield  {author} {\bibinfo {author} {\bibfnamefont {B.~J.}\ \bibnamefont
  {Lawrie}}, \bibinfo {author} {\bibfnamefont {P.~D.}\ \bibnamefont {Lett}},
  \bibinfo {author} {\bibfnamefont {A.~M.}\ \bibnamefont {Marino}},\ and\
  \bibinfo {author} {\bibfnamefont {R.~C.}\ \bibnamefont {Pooser}},\ }\bibfield
   {title} {\enquote {\bibinfo {title} {Quantum sensing with squeezed light},}\
  }\href@noop {} {\bibfield  {journal} {\bibinfo  {journal} {ACS Photonics}\
  }\textbf {\bibinfo {volume} {6}},\ \bibinfo {pages} {1307--1318} (\bibinfo
  {year} {2019})}\BibitemShut {NoStop}%
\bibitem [{\citenamefont {Lee}\ \emph {et~al.}(2021)\citenamefont {Lee},
  \citenamefont {Lawrie}, \citenamefont {Pooser}, \citenamefont {Lee},
  \citenamefont {Rockstuhl},\ and\ \citenamefont {Tame}}]{lee2020quantum}%
  \BibitemOpen
  \bibfield  {author} {\bibinfo {author} {\bibfnamefont {C.}~\bibnamefont
  {Lee}}, \bibinfo {author} {\bibfnamefont {B.}~\bibnamefont {Lawrie}},
  \bibinfo {author} {\bibfnamefont {R.}~\bibnamefont {Pooser}}, \bibinfo
  {author} {\bibfnamefont {K.-G.}\ \bibnamefont {Lee}}, \bibinfo {author}
  {\bibfnamefont {C.}~\bibnamefont {Rockstuhl}},\ and\ \bibinfo {author}
  {\bibfnamefont {M.}~\bibnamefont {Tame}},\ }\bibfield  {title} {\enquote
  {\bibinfo {title} {Quantum plasmonic sensors},}\ }\href@noop {} {\bibfield
  {journal} {\bibinfo  {journal} {Chemical Reviews}\ }\textbf {\bibinfo
  {volume} {121}},\ \bibinfo {pages} {4743--4804} (\bibinfo {year}
  {2021})}\BibitemShut {NoStop}%
\bibitem [{\citenamefont {Aasi}\ \emph {et~al.}(2013)\citenamefont {Aasi},
  \citenamefont {Abadie}, \citenamefont {Abbott}, \citenamefont {Abbott},
  \citenamefont {Abbott}, \citenamefont {Abernathy}, \citenamefont {Adams},
  \citenamefont {Adams}, \citenamefont {Addesso}, \citenamefont {Adhikari}
  \emph {et~al.}}]{aasi2013enhanced}%
  \BibitemOpen
  \bibfield  {author} {\bibinfo {author} {\bibfnamefont {J.}~\bibnamefont
  {Aasi}}, \bibinfo {author} {\bibfnamefont {J.}~\bibnamefont {Abadie}},
  \bibinfo {author} {\bibfnamefont {B.}~\bibnamefont {Abbott}}, \bibinfo
  {author} {\bibfnamefont {R.}~\bibnamefont {Abbott}}, \bibinfo {author}
  {\bibfnamefont {T.}~\bibnamefont {Abbott}}, \bibinfo {author} {\bibfnamefont
  {M.}~\bibnamefont {Abernathy}}, \bibinfo {author} {\bibfnamefont
  {C.}~\bibnamefont {Adams}}, \bibinfo {author} {\bibfnamefont
  {T.}~\bibnamefont {Adams}}, \bibinfo {author} {\bibfnamefont
  {P.}~\bibnamefont {Addesso}}, \bibinfo {author} {\bibfnamefont
  {R.}~\bibnamefont {Adhikari}}, \emph {et~al.},\ }\bibfield  {title} {\enquote
  {\bibinfo {title} {Enhanced sensitivity of the ligo gravitational wave
  detector by using squeezed states of light},}\ }\href@noop {} {\bibfield
  {journal} {\bibinfo  {journal} {Nature Photonics}\ }\textbf {\bibinfo
  {volume} {7}},\ \bibinfo {pages} {613} (\bibinfo {year} {2013})}\BibitemShut
  {NoStop}%
\bibitem [{\citenamefont {Ma}\ \emph {et~al.}(2017)\citenamefont {Ma},
  \citenamefont {Miao}, \citenamefont {Pang}, \citenamefont {Evans},
  \citenamefont {Zhao}, \citenamefont {Harms}, \citenamefont {Schnabel},\ and\
  \citenamefont {Chen}}]{ma2017proposal}%
  \BibitemOpen
  \bibfield  {author} {\bibinfo {author} {\bibfnamefont {Y.}~\bibnamefont
  {Ma}}, \bibinfo {author} {\bibfnamefont {H.}~\bibnamefont {Miao}}, \bibinfo
  {author} {\bibfnamefont {B.~H.}\ \bibnamefont {Pang}}, \bibinfo {author}
  {\bibfnamefont {M.}~\bibnamefont {Evans}}, \bibinfo {author} {\bibfnamefont
  {C.}~\bibnamefont {Zhao}}, \bibinfo {author} {\bibfnamefont {J.}~\bibnamefont
  {Harms}}, \bibinfo {author} {\bibfnamefont {R.}~\bibnamefont {Schnabel}},\
  and\ \bibinfo {author} {\bibfnamefont {Y.}~\bibnamefont {Chen}},\ }\bibfield
  {title} {\enquote {\bibinfo {title} {Proposal for gravitational-wave
  detection beyond the standard quantum limit through epr entanglement},}\
  }\href@noop {} {\bibfield  {journal} {\bibinfo  {journal} {Nature Physics}\
  }\textbf {\bibinfo {volume} {13}},\ \bibinfo {pages} {776} (\bibinfo {year}
  {2017})}\BibitemShut {NoStop}%
\bibitem [{\citenamefont {Fan}, \citenamefont {Lawrie},\ and\ \citenamefont
  {Pooser}(2015)}]{fan2015quantum}%
  \BibitemOpen
  \bibfield  {author} {\bibinfo {author} {\bibfnamefont {W.}~\bibnamefont
  {Fan}}, \bibinfo {author} {\bibfnamefont {B.~J.}\ \bibnamefont {Lawrie}},\
  and\ \bibinfo {author} {\bibfnamefont {R.~C.}\ \bibnamefont {Pooser}},\
  }\bibfield  {title} {\enquote {\bibinfo {title} {Quantum plasmonic
  sensing},}\ }\href@noop {} {\bibfield  {journal} {\bibinfo  {journal}
  {Physical Review A}\ }\textbf {\bibinfo {volume} {92}},\ \bibinfo {pages}
  {053812} (\bibinfo {year} {2015})}\BibitemShut {NoStop}%
\bibitem [{\citenamefont {Dowran}\ \emph {et~al.}(2018)\citenamefont {Dowran},
  \citenamefont {Kumar}, \citenamefont {Lawrie}, \citenamefont {Pooser},\ and\
  \citenamefont {Marino}}]{dowran2018quantum}%
  \BibitemOpen
  \bibfield  {author} {\bibinfo {author} {\bibfnamefont {M.}~\bibnamefont
  {Dowran}}, \bibinfo {author} {\bibfnamefont {A.}~\bibnamefont {Kumar}},
  \bibinfo {author} {\bibfnamefont {B.~J.}\ \bibnamefont {Lawrie}}, \bibinfo
  {author} {\bibfnamefont {R.~C.}\ \bibnamefont {Pooser}},\ and\ \bibinfo
  {author} {\bibfnamefont {A.~M.}\ \bibnamefont {Marino}},\ }\bibfield  {title}
  {\enquote {\bibinfo {title} {Quantum-enhanced plasmonic sensing},}\
  }\href@noop {} {\bibfield  {journal} {\bibinfo  {journal} {Optica}\ }\textbf
  {\bibinfo {volume} {5}},\ \bibinfo {pages} {628--633} (\bibinfo {year}
  {2018})}\BibitemShut {NoStop}%
\bibitem [{\citenamefont {Pooser}\ and\ \citenamefont
  {Lawrie}(2015{\natexlab{a}})}]{pooser_plasmonic_2015}%
  \BibitemOpen
  \bibfield  {author} {\bibinfo {author} {\bibfnamefont {R.~C.}\ \bibnamefont
  {Pooser}}\ and\ \bibinfo {author} {\bibfnamefont {B.}~\bibnamefont
  {Lawrie}},\ }\bibfield  {title} {\enquote {\bibinfo {title} {Plasmonic trace
  sensing below the photon shot noise limit},}\ }\href
  {http://pubs.acs.org/doi/abs/10.1021/acsphotonics.5b00501} {\bibfield
  {journal} {\bibinfo  {journal} {ACS Photonics}\ }\textbf {\bibinfo {volume}
  {3}},\ \bibinfo {pages} {8--13} (\bibinfo {year}
  {2015}{\natexlab{a}})}\BibitemShut {NoStop}%
\bibitem [{\citenamefont {Pooser}\ and\ \citenamefont
  {Lawrie}(2015{\natexlab{b}})}]{pooser_ultrasensitive_2015}%
  \BibitemOpen
  \bibfield  {author} {\bibinfo {author} {\bibfnamefont {R.~C.}\ \bibnamefont
  {Pooser}}\ and\ \bibinfo {author} {\bibfnamefont {B.}~\bibnamefont
  {Lawrie}},\ }\bibfield  {title} {\enquote {\bibinfo {title} {Ultrasensitive
  measurement of microcantilever displacement below the shot-noise limit},}\
  }\href@noop {} {\bibfield  {journal} {\bibinfo  {journal} {Optica}\ }\textbf
  {\bibinfo {volume} {2}},\ \bibinfo {pages} {393--399} (\bibinfo {year}
  {2015}{\natexlab{b}})}\BibitemShut {NoStop}%
\bibitem [{\citenamefont {Pooser}\ \emph {et~al.}(2020)\citenamefont {Pooser},
  \citenamefont {Savino}, \citenamefont {Batson}, \citenamefont {Beckey},
  \citenamefont {Garcia},\ and\ \citenamefont {Lawrie}}]{pooser2020truncated}%
  \BibitemOpen
  \bibfield  {author} {\bibinfo {author} {\bibfnamefont {R.}~\bibnamefont
  {Pooser}}, \bibinfo {author} {\bibfnamefont {N.}~\bibnamefont {Savino}},
  \bibinfo {author} {\bibfnamefont {E.}~\bibnamefont {Batson}}, \bibinfo
  {author} {\bibfnamefont {J.}~\bibnamefont {Beckey}}, \bibinfo {author}
  {\bibfnamefont {J.}~\bibnamefont {Garcia}},\ and\ \bibinfo {author}
  {\bibfnamefont {B.}~\bibnamefont {Lawrie}},\ }\bibfield  {title} {\enquote
  {\bibinfo {title} {Truncated nonlinear interferometry for quantum-enhanced
  atomic force microscopy},}\ }\href@noop {} {\bibfield  {journal} {\bibinfo
  {journal} {Physical Review Letters}\ }\textbf {\bibinfo {volume} {124}},\
  \bibinfo {pages} {230504} (\bibinfo {year} {2020})}\BibitemShut {NoStop}%
\bibitem [{\citenamefont {Otterstrom}, \citenamefont {Pooser},\ and\
  \citenamefont {Lawrie}(2014)}]{otterstrom_nonlinear_2014}%
  \BibitemOpen
  \bibfield  {author} {\bibinfo {author} {\bibfnamefont {N.}~\bibnamefont
  {Otterstrom}}, \bibinfo {author} {\bibfnamefont {R.~C.}\ \bibnamefont
  {Pooser}},\ and\ \bibinfo {author} {\bibfnamefont {B.~J.}\ \bibnamefont
  {Lawrie}},\ }\bibfield  {title} {\enquote {\bibinfo {title} {Nonlinear
  optical magnetometry with accessible in situ optical squeezing},}\ }\href
  {https://www.osapublishing.org/abstract.cfm?uri=ol-39-22-6533} {\bibfield
  {journal} {\bibinfo  {journal} {Optics Letters}\ }\textbf {\bibinfo {volume}
  {39}},\ \bibinfo {pages} {6533--6536} (\bibinfo {year} {2014})}\BibitemShut
  {NoStop}%
\bibitem [{\citenamefont {Wolfgramm}\ \emph {et~al.}(2010)\citenamefont
  {Wolfgramm}, \citenamefont {Cere}, \citenamefont {Beduini}, \citenamefont
  {Predojevi{\'c}}, \citenamefont {Koschorreck},\ and\ \citenamefont
  {Mitchell}}]{wolfgramm2010squeezed}%
  \BibitemOpen
  \bibfield  {author} {\bibinfo {author} {\bibfnamefont {F.}~\bibnamefont
  {Wolfgramm}}, \bibinfo {author} {\bibfnamefont {A.}~\bibnamefont {Cere}},
  \bibinfo {author} {\bibfnamefont {F.~A.}\ \bibnamefont {Beduini}}, \bibinfo
  {author} {\bibfnamefont {A.}~\bibnamefont {Predojevi{\'c}}}, \bibinfo
  {author} {\bibfnamefont {M.}~\bibnamefont {Koschorreck}},\ and\ \bibinfo
  {author} {\bibfnamefont {M.~W.}\ \bibnamefont {Mitchell}},\ }\bibfield
  {title} {\enquote {\bibinfo {title} {Squeezed-light optical magnetometry},}\
  }\href@noop {} {\bibfield  {journal} {\bibinfo  {journal} {Physical Review
  Letters}\ }\textbf {\bibinfo {volume} {105}},\ \bibinfo {pages} {053601}
  (\bibinfo {year} {2010})}\BibitemShut {NoStop}%
\bibitem [{\citenamefont {Li}\ \emph {et~al.}(2018)\citenamefont {Li},
  \citenamefont {Bilek}, \citenamefont {Hoff}, \citenamefont {Madsen},
  \citenamefont {Forstner}, \citenamefont {Prakash}, \citenamefont
  {Sch{\"a}fermeier}, \citenamefont {Gehring}, \citenamefont {Bowen},\ and\
  \citenamefont {Andersen}}]{li2018quantum}%
  \BibitemOpen
  \bibfield  {author} {\bibinfo {author} {\bibfnamefont {B.-B.}\ \bibnamefont
  {Li}}, \bibinfo {author} {\bibfnamefont {J.}~\bibnamefont {Bilek}}, \bibinfo
  {author} {\bibfnamefont {U.~B.}\ \bibnamefont {Hoff}}, \bibinfo {author}
  {\bibfnamefont {L.~S.}\ \bibnamefont {Madsen}}, \bibinfo {author}
  {\bibfnamefont {S.}~\bibnamefont {Forstner}}, \bibinfo {author}
  {\bibfnamefont {V.}~\bibnamefont {Prakash}}, \bibinfo {author} {\bibfnamefont
  {C.}~\bibnamefont {Sch{\"a}fermeier}}, \bibinfo {author} {\bibfnamefont
  {T.}~\bibnamefont {Gehring}}, \bibinfo {author} {\bibfnamefont {W.~P.}\
  \bibnamefont {Bowen}},\ and\ \bibinfo {author} {\bibfnamefont {U.~L.}\
  \bibnamefont {Andersen}},\ }\bibfield  {title} {\enquote {\bibinfo {title}
  {Quantum enhanced optomechanical magnetometry},}\ }\href@noop {} {\bibfield
  {journal} {\bibinfo  {journal} {Optica}\ }\textbf {\bibinfo {volume} {5}},\
  \bibinfo {pages} {850--856} (\bibinfo {year} {2018})}\BibitemShut {NoStop}%
\bibitem [{\citenamefont {Horrom}\ \emph {et~al.}(2012)\citenamefont {Horrom},
  \citenamefont {Singh}, \citenamefont {Dowling},\ and\ \citenamefont
  {Mikhailov}}]{horrom2012quantum}%
  \BibitemOpen
  \bibfield  {author} {\bibinfo {author} {\bibfnamefont {T.}~\bibnamefont
  {Horrom}}, \bibinfo {author} {\bibfnamefont {R.}~\bibnamefont {Singh}},
  \bibinfo {author} {\bibfnamefont {J.~P.}\ \bibnamefont {Dowling}},\ and\
  \bibinfo {author} {\bibfnamefont {E.~E.}\ \bibnamefont {Mikhailov}},\
  }\bibfield  {title} {\enquote {\bibinfo {title} {Quantum-enhanced
  magnetometer with low-frequency squeezing},}\ }\href@noop {} {\bibfield
  {journal} {\bibinfo  {journal} {Physical Review A}\ }\textbf {\bibinfo
  {volume} {86}},\ \bibinfo {pages} {023803} (\bibinfo {year}
  {2012})}\BibitemShut {NoStop}%
\bibitem [{\citenamefont {Michael}\ \emph {et~al.}(2019)\citenamefont
  {Michael}, \citenamefont {Bello}, \citenamefont {Rosenbluh},\ and\
  \citenamefont {Pe’er}}]{michael2019squeezing}%
  \BibitemOpen
  \bibfield  {author} {\bibinfo {author} {\bibfnamefont {Y.}~\bibnamefont
  {Michael}}, \bibinfo {author} {\bibfnamefont {L.}~\bibnamefont {Bello}},
  \bibinfo {author} {\bibfnamefont {M.}~\bibnamefont {Rosenbluh}},\ and\
  \bibinfo {author} {\bibfnamefont {A.}~\bibnamefont {Pe’er}},\ }\bibfield
  {title} {\enquote {\bibinfo {title} {Squeezing-enhanced raman
  spectroscopy},}\ }\href@noop {} {\bibfield  {journal} {\bibinfo  {journal}
  {npj Quantum Information}\ }\textbf {\bibinfo {volume} {5}},\ \bibinfo
  {pages} {1--9} (\bibinfo {year} {2019})}\BibitemShut {NoStop}%
\bibitem [{\citenamefont {de~Andrade}\ \emph {et~al.}(2020)\citenamefont
  {de~Andrade}, \citenamefont {Kerdoncuff}, \citenamefont {Berg-S{\o}rensen},
  \citenamefont {Gehring}, \citenamefont {Lassen},\ and\ \citenamefont
  {Andersen}}]{de2020quantum}%
  \BibitemOpen
  \bibfield  {author} {\bibinfo {author} {\bibfnamefont {R.~B.}\ \bibnamefont
  {de~Andrade}}, \bibinfo {author} {\bibfnamefont {H.}~\bibnamefont
  {Kerdoncuff}}, \bibinfo {author} {\bibfnamefont {K.}~\bibnamefont
  {Berg-S{\o}rensen}}, \bibinfo {author} {\bibfnamefont {T.}~\bibnamefont
  {Gehring}}, \bibinfo {author} {\bibfnamefont {M.}~\bibnamefont {Lassen}},\
  and\ \bibinfo {author} {\bibfnamefont {U.~L.}\ \bibnamefont {Andersen}},\
  }\bibfield  {title} {\enquote {\bibinfo {title} {Quantum-enhanced
  continuous-wave stimulated raman scattering spectroscopy},}\ }\href@noop {}
  {\bibfield  {journal} {\bibinfo  {journal} {Optica}\ }\textbf {\bibinfo
  {volume} {7}},\ \bibinfo {pages} {470--475} (\bibinfo {year}
  {2020})}\BibitemShut {NoStop}%
\bibitem [{\citenamefont {Casacio}\ \emph {et~al.}(2021)\citenamefont
  {Casacio}, \citenamefont {Madsen}, \citenamefont {Terrasson}, \citenamefont
  {Waleed}, \citenamefont {Barnscheidt}, \citenamefont {Hage}, \citenamefont
  {Taylor},\ and\ \citenamefont {Bowen}}]{casacio2021quantum}%
  \BibitemOpen
  \bibfield  {author} {\bibinfo {author} {\bibfnamefont {C.~A.}\ \bibnamefont
  {Casacio}}, \bibinfo {author} {\bibfnamefont {L.~S.}\ \bibnamefont {Madsen}},
  \bibinfo {author} {\bibfnamefont {A.}~\bibnamefont {Terrasson}}, \bibinfo
  {author} {\bibfnamefont {M.}~\bibnamefont {Waleed}}, \bibinfo {author}
  {\bibfnamefont {K.}~\bibnamefont {Barnscheidt}}, \bibinfo {author}
  {\bibfnamefont {B.}~\bibnamefont {Hage}}, \bibinfo {author} {\bibfnamefont
  {M.~A.}\ \bibnamefont {Taylor}},\ and\ \bibinfo {author} {\bibfnamefont
  {W.~P.}\ \bibnamefont {Bowen}},\ }\bibfield  {title} {\enquote {\bibinfo
  {title} {Quantum-enhanced nonlinear microscopy},}\ }\href@noop {} {\bibfield
  {journal} {\bibinfo  {journal} {Nature}\ }\textbf {\bibinfo {volume} {594}},\
  \bibinfo {pages} {201--206} (\bibinfo {year} {2021})}\BibitemShut {NoStop}%
\bibitem [{\citenamefont {Lucivero}\ \emph {et~al.}(2016)\citenamefont
  {Lucivero}, \citenamefont {Jim{\'e}nez-Mart{\'\i}nez}, \citenamefont {Kong},\
  and\ \citenamefont {Mitchell}}]{lucivero2016squeezed}%
  \BibitemOpen
  \bibfield  {author} {\bibinfo {author} {\bibfnamefont {V.~G.}\ \bibnamefont
  {Lucivero}}, \bibinfo {author} {\bibfnamefont {R.}~\bibnamefont
  {Jim{\'e}nez-Mart{\'\i}nez}}, \bibinfo {author} {\bibfnamefont
  {J.}~\bibnamefont {Kong}},\ and\ \bibinfo {author} {\bibfnamefont {M.~W.}\
  \bibnamefont {Mitchell}},\ }\bibfield  {title} {\enquote {\bibinfo {title}
  {Squeezed-light spin noise spectroscopy},}\ }\href@noop {} {\bibfield
  {journal} {\bibinfo  {journal} {Physical Review A}\ }\textbf {\bibinfo
  {volume} {93}},\ \bibinfo {pages} {053802} (\bibinfo {year}
  {2016})}\BibitemShut {NoStop}%
\bibitem [{\citenamefont {Wineland}\ \emph {et~al.}(1992)\citenamefont
  {Wineland}, \citenamefont {Bollinger}, \citenamefont {Itano}, \citenamefont
  {Moore},\ and\ \citenamefont {Heinzen}}]{wineland1992spin}%
  \BibitemOpen
  \bibfield  {author} {\bibinfo {author} {\bibfnamefont {D.~J.}\ \bibnamefont
  {Wineland}}, \bibinfo {author} {\bibfnamefont {J.~J.}\ \bibnamefont
  {Bollinger}}, \bibinfo {author} {\bibfnamefont {W.~M.}\ \bibnamefont
  {Itano}}, \bibinfo {author} {\bibfnamefont {F.}~\bibnamefont {Moore}},\ and\
  \bibinfo {author} {\bibfnamefont {D.}~\bibnamefont {Heinzen}},\ }\bibfield
  {title} {\enquote {\bibinfo {title} {Spin squeezing and reduced quantum noise
  in spectroscopy},}\ }\href@noop {} {\bibfield  {journal} {\bibinfo  {journal}
  {Physical Review A}\ }\textbf {\bibinfo {volume} {46}},\ \bibinfo {pages}
  {R6797} (\bibinfo {year} {1992})}\BibitemShut {NoStop}%
\bibitem [{\citenamefont {Hudelist}\ \emph {et~al.}(2014)\citenamefont
  {Hudelist}, \citenamefont {Kong}, \citenamefont {Liu}, \citenamefont {Jing},
  \citenamefont {Ou},\ and\ \citenamefont {Zhang}}]{OuNatComm}%
  \BibitemOpen
  \bibfield  {author} {\bibinfo {author} {\bibfnamefont {F.}~\bibnamefont
  {Hudelist}}, \bibinfo {author} {\bibfnamefont {J.}~\bibnamefont {Kong}},
  \bibinfo {author} {\bibfnamefont {C.}~\bibnamefont {Liu}}, \bibinfo {author}
  {\bibfnamefont {J.}~\bibnamefont {Jing}}, \bibinfo {author} {\bibfnamefont
  {Z.}~\bibnamefont {Ou}},\ and\ \bibinfo {author} {\bibfnamefont
  {W.}~\bibnamefont {Zhang}},\ }\bibfield  {title} {\enquote {\bibinfo {title}
  {Quantum metrology with parametric amplifier-based photon correlation
  interferometers},}\ }\href {http://dx.doi.org/10.1038/ncomms4049} {\bibfield
  {journal} {\bibinfo  {journal} {Nature Communications}\ }\textbf {\bibinfo
  {volume} {5}},\ \bibinfo {pages} {3049} (\bibinfo {year} {2014})}\BibitemShut
  {NoStop}%
\bibitem [{\citenamefont {Ou}(2012)}]{Ou}%
  \BibitemOpen
  \bibfield  {author} {\bibinfo {author} {\bibfnamefont {Z.~Y.}\ \bibnamefont
  {Ou}},\ }\bibfield  {title} {\enquote {\bibinfo {title} {Enhancement of the
  phase-measurement sensitivity beyond the standard quantum limit by a
  nonlinear interferometer},}\ }\href
  {https://doi.org/10.1103/PhysRevA.85.023815} {\bibfield  {journal} {\bibinfo
  {journal} {Physical Review A}\ }\textbf {\bibinfo {volume} {85}},\ \bibinfo
  {pages} {023815} (\bibinfo {year} {2012})}\BibitemShut {NoStop}%
\bibitem [{\citenamefont {Jing}\ \emph {et~al.}(2011)\citenamefont {Jing},
  \citenamefont {Liu}, \citenamefont {Zhou}, \citenamefont {Ou},\ and\
  \citenamefont {Zhang}}]{jing2011realization}%
  \BibitemOpen
  \bibfield  {author} {\bibinfo {author} {\bibfnamefont {J.}~\bibnamefont
  {Jing}}, \bibinfo {author} {\bibfnamefont {C.}~\bibnamefont {Liu}}, \bibinfo
  {author} {\bibfnamefont {Z.}~\bibnamefont {Zhou}}, \bibinfo {author}
  {\bibfnamefont {Z.}~\bibnamefont {Ou}},\ and\ \bibinfo {author}
  {\bibfnamefont {W.}~\bibnamefont {Zhang}},\ }\bibfield  {title} {\enquote
  {\bibinfo {title} {Realization of a nonlinear interferometer with parametric
  amplifiers},}\ }\href@noop {} {\bibfield  {journal} {\bibinfo  {journal}
  {Applied Physics Letters}\ }\textbf {\bibinfo {volume} {99}},\ \bibinfo
  {pages} {011110} (\bibinfo {year} {2011})}\BibitemShut {NoStop}%
\bibitem [{\citenamefont {Wang}\ \emph {et~al.}(2021)\citenamefont {Wang},
  \citenamefont {Fu}, \citenamefont {Ni}, \citenamefont {Zhang}, \citenamefont
  {Zhao}, \citenamefont {Jin},\ and\ \citenamefont {Jing}}]{wang2021nonlinear}%
  \BibitemOpen
  \bibfield  {author} {\bibinfo {author} {\bibfnamefont {H.}~\bibnamefont
  {Wang}}, \bibinfo {author} {\bibfnamefont {Z.}~\bibnamefont {Fu}}, \bibinfo
  {author} {\bibfnamefont {Z.}~\bibnamefont {Ni}}, \bibinfo {author}
  {\bibfnamefont {X.}~\bibnamefont {Zhang}}, \bibinfo {author} {\bibfnamefont
  {C.}~\bibnamefont {Zhao}}, \bibinfo {author} {\bibfnamefont {S.}~\bibnamefont
  {Jin}},\ and\ \bibinfo {author} {\bibfnamefont {J.}~\bibnamefont {Jing}},\
  }\bibfield  {title} {\enquote {\bibinfo {title} {Nonlinear interferometric
  surface-plasmon-resonance sensor},}\ }\href@noop {} {\bibfield  {journal}
  {\bibinfo  {journal} {Optics Express}\ }\textbf {\bibinfo {volume} {29}},\
  \bibinfo {pages} {11194--11206} (\bibinfo {year} {2021})}\BibitemShut
  {NoStop}%
\bibitem [{\citenamefont {Treps}\ \emph {et~al.}(2003)\citenamefont {Treps},
  \citenamefont {Grosse}, \citenamefont {Bowen}, \citenamefont {Fabre},
  \citenamefont {Bachor},\ and\ \citenamefont {Lam}}]{treps2003quantum}%
  \BibitemOpen
  \bibfield  {author} {\bibinfo {author} {\bibfnamefont {N.}~\bibnamefont
  {Treps}}, \bibinfo {author} {\bibfnamefont {N.}~\bibnamefont {Grosse}},
  \bibinfo {author} {\bibfnamefont {W.~P.}\ \bibnamefont {Bowen}}, \bibinfo
  {author} {\bibfnamefont {C.}~\bibnamefont {Fabre}}, \bibinfo {author}
  {\bibfnamefont {H.-A.}\ \bibnamefont {Bachor}},\ and\ \bibinfo {author}
  {\bibfnamefont {P.~K.}\ \bibnamefont {Lam}},\ }\bibfield  {title} {\enquote
  {\bibinfo {title} {A quantum laser pointer},}\ }\href@noop {} {\bibfield
  {journal} {\bibinfo  {journal} {Science}\ }\textbf {\bibinfo {volume}
  {301}},\ \bibinfo {pages} {940--943} (\bibinfo {year} {2003})}\BibitemShut
  {NoStop}%
\bibitem [{\citenamefont {Anderson}\ \emph
  {et~al.}(2017{\natexlab{a}})\citenamefont {Anderson}, \citenamefont
  {Schmittberger}, \citenamefont {Gupta}, \citenamefont {Jones},\ and\
  \citenamefont {Lett}}]{LettPRA}%
  \BibitemOpen
  \bibfield  {author} {\bibinfo {author} {\bibfnamefont {B.~E.}\ \bibnamefont
  {Anderson}}, \bibinfo {author} {\bibfnamefont {B.~L.}\ \bibnamefont
  {Schmittberger}}, \bibinfo {author} {\bibfnamefont {P.}~\bibnamefont
  {Gupta}}, \bibinfo {author} {\bibfnamefont {K.~M.}\ \bibnamefont {Jones}},\
  and\ \bibinfo {author} {\bibfnamefont {P.~D.}\ \bibnamefont {Lett}},\
  }\bibfield  {title} {\enquote {\bibinfo {title} {Optimal phase measurements
  with bright- and vacuum-seeded su(1,1) interferometers},}\ }\href
  {https://doi.org/10.1103/PhysRevA.95.063843} {\bibfield  {journal} {\bibinfo
  {journal} {Physical Review A}\ }\textbf {\bibinfo {volume} {95}},\ \bibinfo
  {pages} {063843} (\bibinfo {year} {2017}{\natexlab{a}})}\BibitemShut
  {NoStop}%
\bibitem [{\citenamefont {Anderson}\ \emph
  {et~al.}(2017{\natexlab{b}})\citenamefont {Anderson}, \citenamefont {Gupta},
  \citenamefont {Schmittberger}, \citenamefont {Horrom}, \citenamefont
  {Hermann-Avigliano}, \citenamefont {Jones},\ and\ \citenamefont
  {Lett}}]{anderson2017phase}%
  \BibitemOpen
  \bibfield  {author} {\bibinfo {author} {\bibfnamefont {B.~E.}\ \bibnamefont
  {Anderson}}, \bibinfo {author} {\bibfnamefont {P.}~\bibnamefont {Gupta}},
  \bibinfo {author} {\bibfnamefont {B.~L.}\ \bibnamefont {Schmittberger}},
  \bibinfo {author} {\bibfnamefont {T.}~\bibnamefont {Horrom}}, \bibinfo
  {author} {\bibfnamefont {C.}~\bibnamefont {Hermann-Avigliano}}, \bibinfo
  {author} {\bibfnamefont {K.~M.}\ \bibnamefont {Jones}},\ and\ \bibinfo
  {author} {\bibfnamefont {P.~D.}\ \bibnamefont {Lett}},\ }\bibfield  {title}
  {\enquote {\bibinfo {title} {Phase sensing beyond the standard quantum limit
  with a variation on the su (1, 1) interferometer},}\ }\href@noop {}
  {\bibfield  {journal} {\bibinfo  {journal} {Optica}\ }\textbf {\bibinfo
  {volume} {4}},\ \bibinfo {pages} {752--756} (\bibinfo {year}
  {2017}{\natexlab{b}})}\BibitemShut {NoStop}%
\bibitem [{\citenamefont {Gupta}\ \emph {et~al.}(2018)\citenamefont {Gupta},
  \citenamefont {Schmittberger}, \citenamefont {Anderson}, \citenamefont
  {Jones},\ and\ \citenamefont {Lett}}]{gupta2018optimized}%
  \BibitemOpen
  \bibfield  {author} {\bibinfo {author} {\bibfnamefont {P.}~\bibnamefont
  {Gupta}}, \bibinfo {author} {\bibfnamefont {B.~L.}\ \bibnamefont
  {Schmittberger}}, \bibinfo {author} {\bibfnamefont {B.~E.}\ \bibnamefont
  {Anderson}}, \bibinfo {author} {\bibfnamefont {K.~M.}\ \bibnamefont
  {Jones}},\ and\ \bibinfo {author} {\bibfnamefont {P.~D.}\ \bibnamefont
  {Lett}},\ }\bibfield  {title} {\enquote {\bibinfo {title} {Optimized phase
  sensing in a truncated su (1, 1) interferometer},}\ }\href@noop {} {\bibfield
   {journal} {\bibinfo  {journal} {Optics Express}\ }\textbf {\bibinfo {volume}
  {26}},\ \bibinfo {pages} {391--401} (\bibinfo {year} {2018})}\BibitemShut
  {NoStop}%
\bibitem [{\citenamefont {Prajapati}\ and\ \citenamefont
  {Novikova}(2019)}]{prajapati2019polarization}%
  \BibitemOpen
  \bibfield  {author} {\bibinfo {author} {\bibfnamefont {N.}~\bibnamefont
  {Prajapati}}\ and\ \bibinfo {author} {\bibfnamefont {I.}~\bibnamefont
  {Novikova}},\ }\bibfield  {title} {\enquote {\bibinfo {title}
  {Polarization-based truncated su(1,1) interferometer based on four-wave
  mixing in rb vapor},}\ }\href {https://doi.org/10.1364/OL.44.005921}
  {\bibfield  {journal} {\bibinfo  {journal} {Opt. Lett.}\ }\textbf {\bibinfo
  {volume} {44}},\ \bibinfo {pages} {5921--5924} (\bibinfo {year}
  {2019})}\BibitemShut {NoStop}%
\bibitem [{\citenamefont {Simon}\ and\ \citenamefont
  {Mukunda}(1990)}]{SimonMukundaMinimal}%
  \BibitemOpen
  \bibfield  {author} {\bibinfo {author} {\bibfnamefont {R.}~\bibnamefont
  {Simon}}\ and\ \bibinfo {author} {\bibfnamefont {N.}~\bibnamefont
  {Mukunda}},\ }\bibfield  {title} {\enquote {\bibinfo {title} {Minimal
  three-component su(2) gadget for polarization optics},}\ }\href
  {https://doi.org/https://doi.org/10.1016/0375-9601(90)90732-4} {\bibfield
  {journal} {\bibinfo  {journal} {Physics Letters A}\ }\textbf {\bibinfo
  {volume} {143}},\ \bibinfo {pages} {165--169} (\bibinfo {year}
  {1990})}\BibitemShut {NoStop}%
\bibitem [{\citenamefont {Bachor}\ and\ \citenamefont
  {Ralph}(2019)}]{bachor2019guide}%
  \BibitemOpen
  \bibfield  {author} {\bibinfo {author} {\bibfnamefont {H.}~\bibnamefont
  {Bachor}}\ and\ \bibinfo {author} {\bibfnamefont {T.}~\bibnamefont {Ralph}},\
  }\href {https://books.google.com/books?id=gvyhDwAAQBAJ} {\emph {\bibinfo
  {title} {A Guide to Experiments in Quantum Optics}}}\ (\bibinfo  {publisher}
  {Wiley},\ \bibinfo {year} {2019})\BibitemShut {NoStop}%
\bibitem [{\citenamefont {Johansson}\ \emph {et~al.}(2013)\citenamefont
  {Johansson} \emph {et~al.}}]{mpmath}%
  \BibitemOpen
  \bibfield  {author} {\bibinfo {author} {\bibfnamefont {F.}~\bibnamefont
  {Johansson}} \emph {et~al.},\ }\href@noop {} {\emph {\bibinfo {title}
  {mpmath: a {P}ython library for arbitrary-precision floating-point arithmetic
  (version 0.18)}}} (\bibinfo {year} {2013}),\ \bibinfo {note} {{\tt
  http://mpmath.org/}}\BibitemShut {NoStop}%
\bibitem [{\citenamefont {Vahlbruch}\ \emph {et~al.}(2016)\citenamefont
  {Vahlbruch}, \citenamefont {Mehmet}, \citenamefont {Danzmann},\ and\
  \citenamefont {Schnabel}}]{Vahlbruch15dB}%
  \BibitemOpen
  \bibfield  {author} {\bibinfo {author} {\bibfnamefont {H.}~\bibnamefont
  {Vahlbruch}}, \bibinfo {author} {\bibfnamefont {M.}~\bibnamefont {Mehmet}},
  \bibinfo {author} {\bibfnamefont {K.}~\bibnamefont {Danzmann}},\ and\
  \bibinfo {author} {\bibfnamefont {R.}~\bibnamefont {Schnabel}},\ }\bibfield
  {title} {\enquote {\bibinfo {title} {Detection of 15 db squeezed states of
  light and their application for the absolute calibration of photoelectric
  quantum efficiency},}\ }\href
  {https://doi.org/10.1103/PhysRevLett.117.110801} {\bibfield  {journal}
  {\bibinfo  {journal} {Phys. Rev. Lett.}\ }\textbf {\bibinfo {volume} {117}},\
  \bibinfo {pages} {110801} (\bibinfo {year} {2016})}\BibitemShut {NoStop}%
\end{thebibliography}%

\end{document}


\maketitle

\section{Polarization rotation to phase shift transduction}
The polarization rotation due to the birefringence of the sample is fundamentally equivalent to the phase shift between left circularly polarized field and right circularly polarized field. A direct way of transducing polarization rotation to phase shift is using this simple fact: two quarter waveplates (QWP) convert the linearly polarized field to circular, interact with sample, then back. Without loss of generality, using the Jones formalism, with $H$ polarized field in, the output fields after a QWP at $+$45$^{\circ}$, sample, then a QWP at $-$45$^{\circ}$, is given by:    

\begin{equation} \label{eq1}
\begin{split}
\left(
\begin{array}{c}
 H \\
 V \\
\end{array}
\right) = & \left(
\begin{array}{cc}
 \left(\frac{1}{2}+\frac{i}{2}\right) e^{-\frac{1}{4} (i \pi )} & \left(-\frac{1}{2}+\frac{i}{2}\right) e^{-\frac{1}{4} (i \pi )} \\
 \left(-\frac{1}{2}+\frac{i}{2}\right) e^{-\frac{1}{4} (i \pi )} & \left(\frac{1}{2}+\frac{i}{2}\right) e^{-\frac{1}{4} (i \pi )} \\
\end{array}
\right) \cdot \left(
\begin{array}{cc}
 \cos \left(\theta _F\right) & \sin \left(\theta _F\right) \\
 -\sin \left(\theta _F\right) & \cos \left(\theta _F\right) \\
\end{array}
\right) \\
& \cdot \left(
\begin{array}{cc}
 \left(\frac{1}{2}+\frac{i}{2}\right) e^{-\frac{1}{4} (i \pi )} & \left(\frac{1}{2}-\frac{i}{2}\right) e^{-\frac{1}{4} (i \pi )} \\
 \left(\frac{1}{2}-\frac{i}{2}\right) e^{-\frac{1}{4} (i \pi )} & \left(\frac{1}{2}+\frac{i}{2}\right) e^{-\frac{1}{4} (i \pi )} \\
\end{array}
\right) \cdot \left(
\begin{array}{c}
 1 \\
 0 \\
\end{array}
\right) \\ 
= & \left(
\begin{array}{c}
 \cos \left(\theta _F\right)-i \sin \left(\theta _F\right) \\
 0 \\
\end{array}
\right)
\end{split}
\end{equation}
This results in a phase shift: 
\begin{equation}
\phi = arg(\text{cos}(\theta_F) - i\text{sin}(\theta_F))
\label{eq:refname1} \end{equation}
\section{Classical interferometer}

\begin{figure}[htbp]
\centering
\includegraphics[width=.6\linewidth]{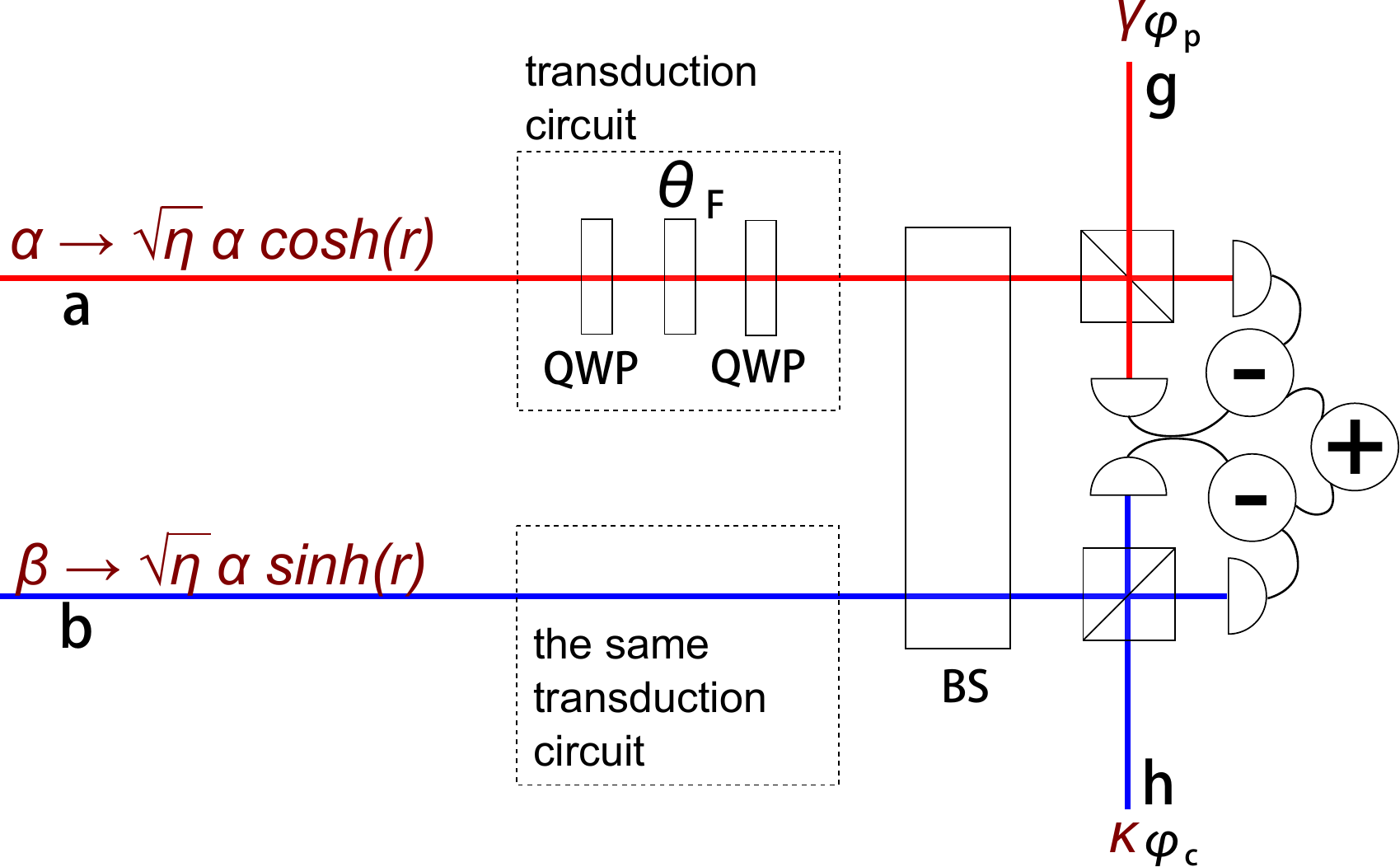}
\caption{The classical interferometer for the quantum enhanced interferometer to benchmark against. }
\label{fig:class}
\end{figure}

Figure \ref{fig:class} is the optics train of a classical interferometer used for benchmarking the quantum enhanced SU(1, 1) truncated interferometer (tSU(1, 1)). The input field a, b are sent to the sample through the polarization-to-phase transduction circuit. The two beams then interact via a beamsplitter (BS). The two output ports of the beamsplitter are readout by homodyne detection with local oscillator (LO) fields g and h. The eigenvalues for the fields a, b, g, and h are $\alpha$, $\beta$, $\gamma$, and $\kappa$. The eigenvalues $\alpha$ and $\beta$ are then rescaled by $\alpha \to \alpha  \sqrt{\eta } \cosh (r)$, $\beta \to \alpha  \sqrt{\eta } \sinh (r)$ such that the number of photons interacting with the sample on each arm is the same as the tSU(1, 1) case, and the photon number for each LO in the classical case is the same as the photon number of each LO in the tSU(1, 1).

\setlength{\parskip}{1em}

{\setlength{\parindent}{0cm} To start, the optical fields of the two inputs of the probe arm homodyne detector can be written by: } 

\begin{equation}
a_f=\frac{a e^{i \phi }+i g e^{i \phi _p}}{\sqrt{2}}  \label{eq:refname1} \end{equation}

\begin{equation} a_n=\frac{i a e^{i \phi }+g e^{i \phi _p}}{\sqrt{2}} 
\label{eq:refname1} \end{equation}

{\setlength{\parindent}{0cm} and the two inputs for the conjugate arm homodyne detector can be written by: } 

\begin{equation} b_f=\frac{b e^{i \phi }+i h e^{i \phi _c}}{\sqrt{2}}
\label{eq:refname1} \end{equation}

\begin{equation} b_n=\frac{i b e^{i \phi }+h e^{i \phi _c}}{\sqrt{2}}
\label{eq:refname1} \end{equation}

{\setlength{\parindent}{0cm} The resulting joint rotated quadrature detection operator J and its expectation value: } 

\begin{equation}
J = a_f^{\dagger}\cdot a_f-  a_n^{\dagger}\cdot a_n  + b_f^{\dagger}\cdot b_f-a_n^{\dagger}\cdot a_n
\label{eq:refname1} \end{equation}

\begin{equation}
\langle J \rangle  = \left\langle \alpha _{\hat{a}},\beta _{\hat{b}},0_{\hat{c}},0_{\hat{d}},0_{\hat{e}},0_{\hat{f}},\gamma _{\hat{g}},\kappa _{\hat{h}}\right|  J  \left|\alpha _{\hat{a}},\beta _{\hat{b}},0_{\hat{c}},0_{\hat{d}},0_{\hat{e}},0_{\hat{f}},\gamma _{\hat{g}},\kappa _{\hat{h}}\right\rangle
\label{eq:refname1} \end{equation}

\begin{equation}
\langle J \rangle  = i e^{-i \left(\phi _c+\phi _p+\phi \right)} \left(-\alpha  \gamma  e^{i \left(\phi _c+2 \phi \right)}+\alpha  \gamma  e^{i \left(\phi _c+2 \phi _p\right)}+\beta  \kappa  e^{i \left(2 \phi _c+\phi _p\right)}-\beta  \kappa  e^{i \left(\phi _p+2 \phi \right)}\right)
\label{eq:refname1} \end{equation}

{\setlength{\parindent}{0cm} The uncertainty: } 
\begin{equation}
\Delta ^2 J = \langle J^2 \rangle  - \langle J \rangle^2 = \alpha ^2+\beta ^2+\gamma ^2+\kappa ^2
\label{eq:refname1} \end{equation}

{\setlength{\parindent}{0cm} The limit of detection (LOD): } 

\begin{equation}
\text{LOD} = \sqrt{\Delta ^2 \phi } =  \sqrt{\frac{\Delta ^2 J}{\left|\langle \frac{\partial J}{\partial \theta _{\theta_F}}\rangle \right| ^2}} = \sqrt{\frac{\Delta ^2 J}{\left| \langle \frac{\partial J}{\partial \phi }\rangle \right| ^2 }}
\label{eq:refname1} \end{equation}
The LOD uses the fact that the transduction from polarization rotation to phase shift is $+1$ or $-1$, depending on the angles of the QWPs. 

{\setlength{\parindent}{0cm} When rescaling $\alpha$ and $\beta$ such that the number of the photons interacting with the sample is the same as those in tSU(1, 1): 
\begin{equation}
\alpha \to \alpha  \sqrt{\eta } \cosh (r),\;  \beta \to \alpha  \sqrt{\eta } \sinh (r)
\label{eq:refname1} \end{equation}

{\setlength{\parindent}{0cm} one obtains: } 

\begin{equation}
\Delta ^2 J = (\gamma ^2+\kappa ^2+\alpha ^2 \eta  \sinh ^2(r)+\alpha ^2 \eta  \cosh ^2(r))
\label{eq:refname1} \end{equation}

\begin{equation} \label{eq1}
\begin{split}
\left| \langle \frac{\partial J}{\partial \theta _{\mathbf{f}}} \rangle \right| ^2  = & \big( \alpha  \gamma  \sqrt{\eta } \cosh (r) e^{i \left(\phi _c+2 \phi _p\right)}+\alpha  \sqrt{\eta } \kappa  \sinh (r) e^{i \left(2 \phi _c+\phi _p\right)}  \\
 + & \alpha  \gamma  \sqrt{\eta } e^{i \left(\phi _c+2 \phi \right)} \cosh (r) + \alpha  \sqrt{\eta } \kappa  e^{i \left(\phi _p+2 \phi \right)} \sinh (r) \big)^* \\ 
 \times &  \big( \alpha  \gamma  \sqrt{\eta } \cosh (r) e^{i \left(\phi _c+2 \phi _p\right)} +\alpha  \sqrt{\eta } \kappa  \sinh (r) e^{i \left(2 \phi _c+\phi _p\right)}\\
 +&\alpha  \gamma  \sqrt{\eta } e^{i \left(\phi _c+2 \phi \right)} \cosh (r)+\alpha  \sqrt{\eta } \kappa  e^{i \left(\phi _p+2 \phi \right)} \sinh (r)\big)
\end{split}
\end{equation}

{\setlength{\parindent}{0cm} Now, with the values below, the LOD is -68.3369 dB (rad): } 

\begin{equation}
r\to 0.88,\;\alpha \to 2\times 10^6,\;\eta \to 1.0,\;\theta \to 0,\;\phi \to 0.001,\;\gamma \to 2\times 10^8,\;\kappa \to 2\times 10^8
\label{eq:refname1} \end{equation}

{\setlength{\parindent}{0cm} and the LOD is -67.8524 dB (rad) with $\eta  = 0.8 $: } 

\begin{equation}
r\to 0.88,\;\alpha \to 2\times 10^6,\;\eta \to 0.8,\;\theta \to 0,\;\phi \to 0.001,\;\gamma \to 2\times 10^8,\;\kappa \to 2\times 10^8
\label{eq:refname1} \end{equation}

\section{tSU(1, 1)}
Figure \ref{fig:tsu11} is the optics train of a full SU(1, 1) interferometer where both the probe (red) and the conjugate (blue) arm are used for polarization rotation sensing. The input fields a (coherent state eigenvalue $\alpha$) and b seed the first optical parametric amplifier (OPA) (squeezing parameter $r$). The resulting two squeezed modes u and v are sent through the transduction circuit. The two modes are then coupled to internal loss   $1 - \eta_{p1}$ and $1 - \eta_{c1}$ through beamsplitters with vacuum fields c and d. The resulting two modes w and z are then sent into a second OPA, and become x and y. They are then coupled to external loss $1 - \eta_{p2}$ and $ 1 -\eta_{c2}$,  and vacuum fields e and f. The final fields before the homodyne detectors are m and n. The LOs have fields g (eigenvalue $\gamma$) and h (eigenvalue $\kappa$).  

\begin{figure}[htbp]
\centering
\includegraphics[width=.6\linewidth]{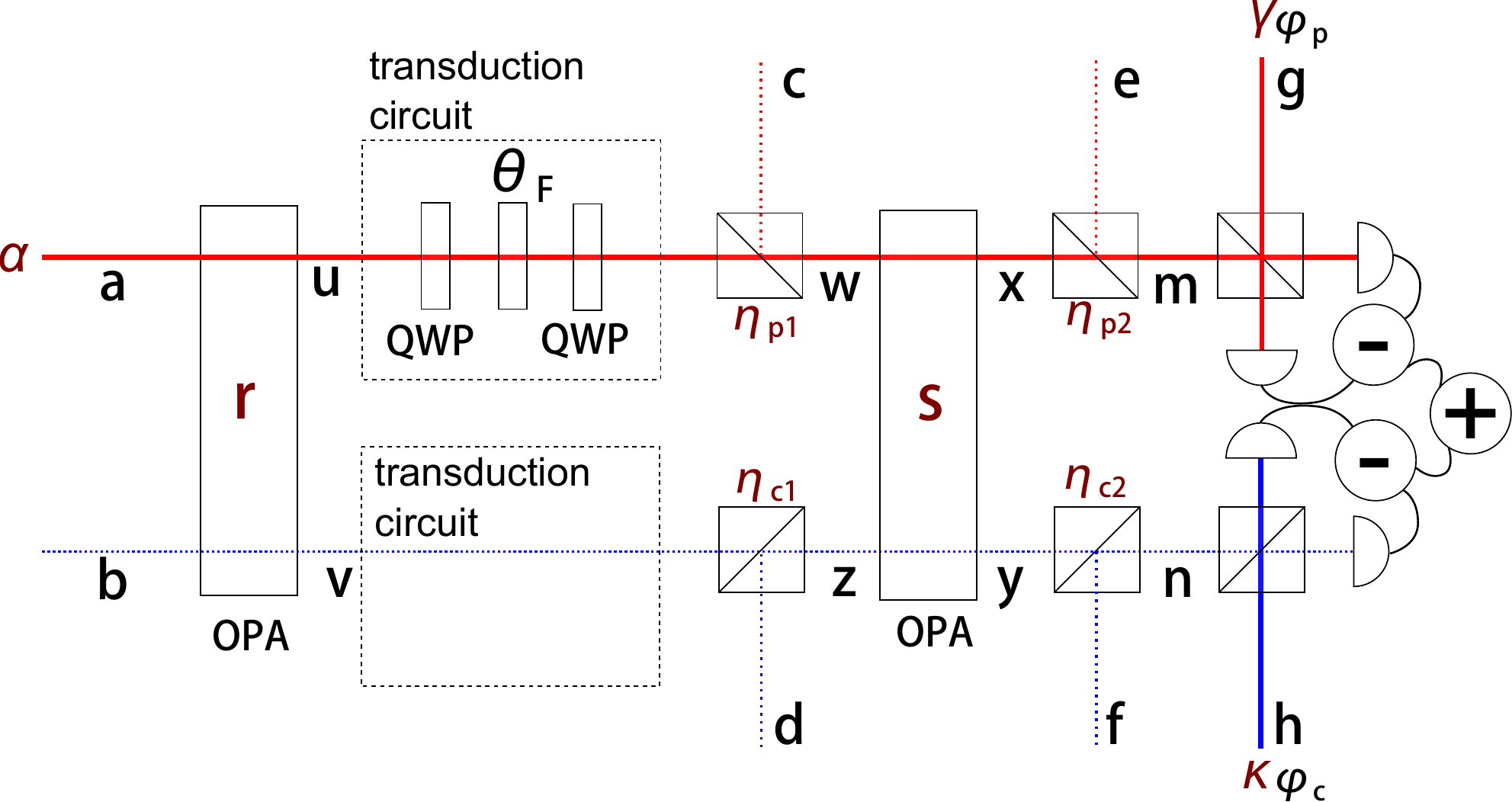}
\caption{Schematics for the full SU(1, 1) considered in the main text. }
\label{fig:tsu11}
\end{figure}

{\setlength{\parindent}{0cm} The optical fields in various points of the optics train can be written as: } 

\begin{equation} u=a \cosh (r)+b^{\dagger } \sinh (r)           \label{eq:refname1} \end{equation}
\begin{equation} v=a^{\dagger } \sinh (r)+b \cosh (r)           \label{eq:refname1} \end{equation}
\begin{equation}     w=i c \sqrt{1-\eta _{\text{p1}}}+u e^{i \phi } \sqrt{\eta _{\text{p1}}}       \label{eq:refname1} \end{equation}
\begin{equation}     z=i d \sqrt{1-\eta _{\text{c1}}}+v e^{i \phi } \sqrt{\eta _{\text{c1}}}       \label{eq:refname1} \end{equation}

\begin{equation}     x=w \cosh (s)+\sinh (s) z^{\dagger }       \label{eq:refname1} \end{equation}
\begin{equation}     y=\sinh (s) w^{\dagger }+z \cosh (s)      \label{eq:refname1} \end{equation}

\begin{equation}     m=x \sqrt{\eta _{\text{p2}}}+i e \sqrt{1-\eta _{\text{p2}}}      \label{eq:refname1} \end{equation}
\begin{equation}     n=y \sqrt{\eta _{\text{c2}}}+i f \sqrt{1-\eta _{\text{c2}}}     \label{eq:refname1} \end{equation}

{\setlength{\parindent}{0cm} The optical fields at the detectors of the homodyne detection can be written as: } 

\begin{equation}     a_f=m \sqrt{\eta _{\text{p3}}}+i g e^{+i \phi _p } \sqrt{1-\eta _{\text{p3}}}       \label{eq:refname1} \end{equation}
\begin{equation}     a_n=g e^{+i \phi _p } \sqrt{\eta _{\text{p3}}}+i m \sqrt{1-\eta _{\text{p3}}}      \label{eq:refname1} \end{equation}

\begin{equation}     b_f=n \sqrt{\eta _{\text{c3}}}+i h e^{+i \phi _c } \sqrt{1-\eta _{\text{c3}}}      \label{eq:refname1} \end{equation}
\begin{equation}     b_n=h e^{+i \phi _c } \sqrt{\eta _{\text{c3}}}+i n \sqrt{1-\eta _{\text{c3}}}      \label{eq:refname1} \end{equation}

{\setlength{\parindent}{0cm} The joint rotated quadrature detection operator J: } 

\begin{equation}
J = a_f^{\dagger}\cdot a_f-a_n^{\dagger}\cdot a_n+b_f^{\dagger}\cdot b_f-b_n^{\dagger}\cdot b_n
\label{eq:refname1} \end{equation}

{\setlength{\parindent}{0cm} The phase shift transduced from polarization rotation: } 
\begin{equation}
\phi = arg(\text{cos}(\theta_F) - i\text{sin}(\theta_F))
\label{eq:refname1} \end{equation}

{\setlength{\parindent}{0cm} which is equivalent to the expression below, which is used to carry out the numerical calculation. } 

\begin{equation}   \phi =  -\frac{1}{2} i \left(\log (\cos (\theta_F)-i \sin (\theta_F))-\log (\cos (\theta_F)-i \sin (\theta_F))^*\right)     \label{eq:refname1} \end{equation}

{\setlength{\parindent}{0cm} To continue, the following substitutions and simplifications are made: $s\to 0$ (from SU(1, 1) to tSU(1, 1)). $\eta _{\text{p1}}\to \eta$ ,$\eta _{\text{c1}}\to \eta$ ,$\eta _{\text{p2}}\to 1$, $\eta _{\text{c2}}\to 1$ (now all the losses are considered \textit{internal}). $\eta _{\text{p3}}\to \frac{1}{2}$, $\eta _{\text{c3}}\to \frac{1}{2}$ for using a regular beam splitter at the homodyne detectors. The resulting J becomes: } 

\begin{equation} \label{eq1}
\begin{split}
J = - &  \left(\frac{e^{-i \phi _c{}} h^{\dagger }}{\sqrt{2}}-\frac{i \left(\sqrt{\eta } e^{-i \phi } \left(a \sinh (r)+b^{\dagger } \cosh (r)\right)-i \sqrt{1-\eta } d^{\dagger }\right)}{\sqrt{2}}\right) \\
\times  &  \left( \frac{i \left(\sqrt{\eta } e^{i \phi } \left(a^{\dagger } \sinh (r)+b \cosh (r)\right)+i d \sqrt{1-\eta }\right)}{\sqrt{2}}+\frac{h e^{i \phi _c}}{\sqrt{2}} \right) \\
+  & \left( \frac{\sqrt{\eta } e^{-i \phi } \left(a \sinh (r)+b^{\dagger } \cosh (r)\right)-i \sqrt{1-\eta } d^{\dagger }}{\sqrt{2}}-\frac{i e^{-i \phi _c{}} h^{\dagger }}{\sqrt{2}} \right) \\
\times  &  \left( \frac{\sqrt{\eta } e^{i \phi } \left(a^{\dagger } \sinh (r)+b \cosh (r)\right)+i d \sqrt{1-\eta }}{\sqrt{2}}+\frac{i h e^{i \phi _c}}{\sqrt{2}} \right) \\
-  & \left(      \frac{g^{\dagger } e^{-i \phi _p{}}}{\sqrt{2}}-\frac{i \left(\sqrt{\eta } e^{-i \phi } \left(a^{\dagger } \cosh (r)+b \sinh (r)\right)-i \sqrt{1-\eta } c^{\dagger }\right)}{\sqrt{2}}     \right) \\
\times  &  \left(    \frac{i \left(\sqrt{\eta } e^{i \phi } \left(a \cosh (r)+b^{\dagger } \sinh (r)\right)+i c \sqrt{1-\eta }\right)}{\sqrt{2}}+\frac{g e^{i \phi _p}}{\sqrt{2}}     \right) \\
+ & \left(     \frac{\sqrt{\eta } e^{-i \phi } \left(a^{\dagger } \cosh (r)+b \sinh (r)\right)-i \sqrt{1-\eta } c^{\dagger }}{\sqrt{2}}-\frac{i g^{\dagger } e^{-i \phi _p{}}}{\sqrt{2}}     \right) \\
\times  &  \left(    \frac{\sqrt{\eta } e^{i \phi } \left(a \cosh (r)+b^{\dagger } \sinh (r)\right)+i c \sqrt{1-\eta }}{\sqrt{2}}+\frac{i g e^{i \phi _p}}{\sqrt{2}}   \right) \\
\end{split} 
\end{equation}

\begin{equation}
\begin{split}
 = & i \sqrt{\eta } \sinh (r) e^{i \phi_c -i \phi } h\cdot a-i \sqrt{\eta } \cosh (r) e^{i \phi -i \phi_p } g^{\dagger }\cdot a\\
+& i \sqrt{\eta } \sinh (r) e^{i \phi_p -i \phi } g\cdot b-i \sqrt{\eta } \cosh (r) e^{i \phi -i \phi_c } h^{\dagger }\cdot b\\ +& i \sqrt{\eta } \sinh (r) e^{i \phi_p -i \phi } g\cdot b-i \sqrt{\eta } \cosh (r) e^{i \phi -i \phi_c } h^{\dagger }\cdot b \\
+& \sqrt{1-\eta } e^{-i \phi_p } g^{\dagger }\cdot c+\sqrt{1-\eta } e^{-i \phi_c } h^{\dagger }\cdot d\\
+& i \sqrt{\eta } \cosh (r) e^{i \phi_p -i \phi } a^{\dagger }\cdot g+\sqrt{1-\eta } e^{i \phi_p } c^{\dagger }\cdot g\\
+& i \sqrt{\eta } \cosh (r) e^{i \phi_c -i \phi } b^{\dagger }\cdot h+\sqrt{1-\eta } e^{i \phi_c } d^{\dagger }\cdot h\\
-& i \sqrt{\eta } \sinh (r) e^{i \phi -i \phi_c } h^{\dagger }\cdot a^{\dagger }-i \sqrt{\eta } \sinh (r) e^{i \phi -i \phi_p } g^{\dagger }\cdot b^{\dagger }\\
\end{split}
\end{equation}

{\setlength{\parindent}{0cm}and the resulting expectation value becomes: } 

\begin{equation} \label{eq1}
\begin{split}
\langle J \rangle = i \alpha  \sqrt{\eta } e^{-i (\phi +\phi _p +\phi _c )} \left(\gamma  \left(-e^{i \phi _c }\right) \cosh (r) \left(e^{2 i \phi }-e^{2 i \phi _p }\right)-\kappa  e^{i \phi _p } \sinh (r) \left(e^{2 i \phi }-e^{2 i \phi _c }\right)\right)\\
\end{split} 
\end{equation}

\begin{equation} \label{eq1}
\begin{split}
\left| \frac{\partial \langle j\rangle }{\partial \phi }\right| ^2 = & \alpha ^2 \eta  \left(\gamma  e^{i \phi _c } \cosh (r) \left(e^{2 i \phi }+e^{2 i \phi _p }\right)+\kappa  e^{i \phi _p } \sinh (r) \left(e^{2 i \phi }+e^{2 i \phi _c }\right)\right)^* \\
\times & \left(\gamma  e^{i \phi _c } \cosh (r) \left(e^{2 i \phi }+e^{2 i \phi _p }\right)+\kappa  e^{i \phi _p } \sinh (r) \left(e^{2 i \phi }+e^{2 i \phi _c }\right)\right)\\
\end{split} 
\end{equation}

\begin{equation} \label{eq1}
\begin{split}
\langle j^2\rangle -\langle j\rangle ^2 =& -\gamma ^2 \eta +\gamma ^2-\eta  \kappa ^2+\kappa ^2+\alpha ^2 \gamma ^2 \eta  \cosh ^2(r) \left(-e^{-2 i (\phi +\phi _p )}\right) \left(e^{2 i \phi }-e^{2 i \phi _p }\right)^2  \\
\times & \alpha ^2 \gamma ^2 \eta  \cosh ^2(r) + \alpha ^2 \eta  \cosh (2 r)+\gamma ^2 \eta  \cosh (2 r)+\eta  \kappa ^2 \cosh (2 r)\\
+ & \alpha ^2 \eta  \kappa ^2 \sinh ^2(r) \left(-e^{-2 i (\phi -\phi _c )}\right)+2 \alpha ^2 \eta  \kappa ^2 \sinh ^2(r)+2 \eta  \sinh ^2(r)\\
- & \alpha ^2 \eta  \kappa ^2 \sinh ^2(r) e^{2 i (\phi -\phi _c )} + \alpha ^2 \eta\\
\times & \left(\gamma  e^{i \phi _c } \cosh (r) \left(e^{2 i \phi }-e^{2 i \phi _p }\right)+\kappa  e^{i \phi _p } \sinh (r) \left(e^{2 i \phi }-e^{2 i \phi _c }\right)\right)^*\\
\times & \gamma  \left(-e^{i \phi _c }\right) \cosh (r) \left(e^{2 i \phi }-e^{2 i \phi _p }\right)-\kappa  e^{i \phi _p } \sinh (r) \left(e^{2 i \phi }-e^{2 i \phi _c }\right)\\
-&\gamma  \eta  \kappa  \sinh (2 r) e^{i (2 \phi -\phi _p -\phi _c )}-\gamma  \eta  \kappa  \sinh (2 r) e^{-2 i \phi +i \phi _p +i \phi _c }\\
-&\alpha ^2 \gamma  \eta  \kappa  \sinh (2 r) e^{i (2 \phi -\phi _p -\phi _c )}+\alpha ^2 \gamma  \eta  \kappa  \sinh (2 r) e^{-i (\phi _p -\phi _c )}\\
+&\alpha ^2 \gamma  \eta  \kappa  \sinh (2 r) e^{i (\phi _p -\phi _c )}-\alpha ^2 \gamma  \eta  \kappa  \sinh (2 r) e^{-2 i \phi +i \phi _p +i \phi _c }\\
\end{split} 
\end{equation}

{\setlength{\parindent}{0cm}and the resulting LOD becomes: } 

\begin{equation}
\text{LOD} = \sqrt{\Delta ^2 \phi } =  \sqrt{\frac{\Delta ^2 J}{\left|\langle \frac{\partial J}{\partial \theta _{\theta_F}}\rangle \right| ^2}} = \sqrt{\frac{\Delta ^2 J}{\left| \langle \frac{\partial J}{\partial \phi }\rangle \right| ^2 }}
\label{eq:refname1} \end{equation}

LOD$_{\text{tSU(1, 1)}}$ - LOD$_{\text{classical}}$  = - 3.854 dB with the parameters below: 
\begin{equation}
r\to 0.88,\;\alpha \to 2\times 10^6,\;\eta \to 1.0,\;\theta \to 0,\;\phi \to 0.001,\;\gamma \to 2\times 10^8,\;\kappa \to 2\times 10^8
\label{eq:refname1} \end{equation}
 and it equals -2.374 dB with the same parameters, except with $\eta = 0.8$.

\section{tSU(1, 1) with vacuum seeding}
One fundamental difference between sensing with a classical light field and a squeezed light field is that the uncertainty of a measurement in the latter has a nontrivial dependence on the phase shift imparted within the interferometer. To see this, with two unseeded input fields, the expectation value of the rotated quadrature operator vanishes:  

\begin{equation} \langle J\rangle  = \left\langle 0_{\hat{a}},0_{\hat{b}},0_{\hat{c}},0_{\hat{d}},0_{\hat{e}},0_{\hat{f}},\gamma _{\hat{g}},\kappa _{\hat{h}}\right| J  \left|0_{\hat{a}},0_{\hat{b}},0_{\hat{c}},0_{\hat{d}},0_{\hat{e}},0_{\hat{f}},\gamma _{\hat{g}},\kappa _{\hat{h}}\right\rangle = 0        \label{eq:refname1} \end{equation}

{\setlength{\parindent}{0cm} and so does the partial derivative: } 

\begin{equation}
\left| \frac{\partial \langle j\rangle }{\partial \phi }\right| ^2 = 0
 \label{eq:refname1} \end{equation}

{\setlength{\parindent}{0cm} The uncertainty, however, depends on $\phi$:} 

\begin{equation} \label{eq1}
\begin{split}
\langle j\rangle ^2 =& 2 \eta  \sinh (r) \left(\sinh (r)-\gamma  \kappa  \cosh (r) e^{-2 i \phi +i \phi_p+i \phi_c }\right) \\ 
+ & \gamma  \left(\gamma  (-\eta )+\gamma + \gamma  \eta  \cosh (2 r) -\eta  \kappa  \sinh (2 r) e^{i (2 \phi -\phi_p-\phi_c )}\right) \\
+ & \kappa ^2 (-\eta +\eta  \cosh (2 r)+1)\\
\end{split} 
\end{equation}